\documentclass[superscriptaddress,onecolumn]{revtex4}
\usepackage{graphicx} 
\usepackage{tikz-cd}
\usepackage{amsmath}
\usepackage{amsfonts}
\usepackage{amssymb}
\usepackage{adjustbox}
\usepackage{appendix}
\usepackage{xcolor}

\begin{document}
\title{
Defect Kinematics in 2D Nematics: Contributions from Surface Topology, Intrinsic and Extrinsic Geometry, Solitons, Defect Orientations, and Elastic Anisotropy 
}
\author{Joseph Pollard}
\email{joe.pollard@unsw.edu.au}
\affiliation{School of Physics, UNSW, Sydney, NSW 2052, Australia.}
\affiliation{EMBL Australia Node in Single Molecule Science, School of Medical Sciences, UNSW, Sydney, NSW 2052, Australia.}
\author{Richard G.~Morris}
\email{r.g.morris@unsw.edu.au}
\affiliation{School of Physics, UNSW, Sydney, NSW 2052, Australia.}
\affiliation{EMBL Australia Node in Single Molecule Science, School of Medical Sciences, UNSW, Sydney, NSW 2052, Australia.}
\affiliation{ARC Centre of Excellence for the Mathematical Analysis of Cellular Systems, UNSW Node, Sydney, NSW 2052, Australia.}
\begin{abstract}
    We characterise the particlelike kinematics of charge-carrying topological defects in nematic media via a geometric field theory.
    This differs from the theory of electromagnetism, with which it is often compared, due to the absence of gauge-invariance.
    In both approaches, basic defect interactions are governed by a propagator, which depends upon the global topology and/or intrinsic geometry of the surface.
    For nematic materials, however, the minimisation of the free energy is sensitive to constraints that a gauge invariant theory would otherwise be indifferent to.
    Hodge theory is used to capture these as `harmonic' excitations, unifying two factors known to additionally affect the kinematics of defects in nematics: relative defect orientations and topological solitons.
    Perturbations to the form of the energy are also permitted in nematic materials due to gauge \emph{non}invariance.
    Those that introduce non-linearities in the corresponding Euler--Lagrange equations are shown to result in defect interactions that go beyond pairwise despite the otherwise abelian nature of the underlying $U(1)$ symmetry.
    We show how this type of induced many-body effect manifests in the cases of non-zero extrinsic curvature and/or elastic anisotropy.
\end{abstract}

\maketitle

\section{Introduction}
The close analogy between nematic materials and electromagnetism is exemplified by the fact that topological defects in two-dimensional nematics interact logarithmically via a pairwise, Coulomb force. This correspondence motivates `particlelike' descriptions of nematic media that centre on the kinematics of charge-carrying topological defects---so-called Coulomb gases~\cite{bowick_symmetry_2022}. Such an approach has been applied in the context of both active and passive liquid crystals~\cite{degennesPhysicsLiquidCrystals2013, vromansOrientationalPropertiesNematic2016, tangOrientationTopologicalDefects2017, angheluta_role_2021, schimming_defect_2025, romano_dynamical_2023, romano_dynamical_2024} and the XY-model~\cite{kosterlitzOrderingMetastabilityPhase1973, vitelli_anomalous_2004, vitelli_defect_2004, turner_vortices_2010}, as well as in related models more broadly~\cite{merminTopologicalTheoryDefects1979}.  

Despite this apparent duality, however, nematic materials are not equivalent to electromagnetism in 2D. 
The difference relates to how the otherwise shared $U(1)$ symmetry manifests in each case. In electromagnetism, such a symmetry underpins a (Yang-Mills) gauge field theory, where the energy is invariant under generic gauge transformations valued in the group $U(1)$. In nematic materials, however, the Frank free energy is manifestly \emph{not} invariant under such a transformation, resulting in a so-called \emph{geometric} field theory where quantities that would be irrelevant in electromagnetism play an important role.
 
Here, we construct a geometric field theory for nematic media in two-dimensions using the modern language of differential forms and exterior calculus. As part of this, we highlight the topological nature of the decomposition of the nematic field that comes from the Hodge theorem~\cite{hodge_theory_1989}. Broadly, this says that a nematic field extremising the energy is generated by three contributions: topologically-nontrivial fields that encode defect winding; a different kind of topologically-nontrivial field that encodes solitons, and; topologically-trivial fields that encode transient geometric features, such as defect orientation, which are not conserved by the dynamics but still play a role in the defect-defect interaction. The possibility of including the latter two parts is where the physics of nematic materials deviate strongly from electromagentism: in a gauge invariant theory, these contributions can be removed by gauge transformations and thus are non-physical, and do not manifest in the energy. Via explicit calculations, our approach leads to two broad contributions.

First, the geometry-centric framework provides a clarifying perspective on many of the effects that have been reported in the literature so far~\cite{vitelli_anomalous_2004, vitelli_defect_2004, turner_vortices_2010, vromansOrientationalPropertiesNematic2016, tangOrientationTopologicalDefects2017, copar_manydefect_2024, schimming_defect_2025}. Capturing the pairwise interactions between defects via a propagator, or Green's function, we describe how such interactions are modified by surface topology due to the existence of non-trivial cycles, as well as intrinsic geometry, due to altered geodesic distances~\cite{vitelli_anomalous_2004, vitelli_defect_2004, turner_vortices_2010}. This corresponds to the first type of field described above, and is therefore analogous to the theory of electromagnetism. There is also an additional affect due to intrinsic geometry---not strictly related to the aforementioned decomposition---with defects experiencing an effective potential whose leading-order contribution is proportional to Gaussian curvature~\cite{vitelli_anomalous_2004, vitelli_defect_2004, turner_vortices_2010}. The other two types of `extra' field in the Hodge decomposition do not play a role in gauge-invariant theories such as electromagnetism. They exert forces on defects without modifying the propagator.
Topological solitons~\cite{schimming_defect_2025} are shown to give rise to such a field, which does not impact defect interactions but can lead to defect nucleation. Further contributions are shown to result from the relative orientations between defects~\cite{vromansOrientationalPropertiesNematic2016, tangOrientationTopologicalDefects2017, pearceOrientationalCorrelationsActive2021,copar_manydefect_2024}; we clarify how such orientational effects manifest on a curved surface, in contrast to previous work that has focussed on flat surfaces. 

Second, the approach also makes transparent a further consequence of geometric field theory: because the energy functional is not required to be gauge invariant, various types of perturbation are allowed that would otherwise be forbidden. Of particular interest are those that render the corresponding Euler--Lagrange equations nonlinear. These perturbations, we show, generically result in \emph{many-body} interactions, going beyond the purely pairwise interactions that occur in the linear systems. This subverts expectations based on naive analogy with electromagentism, since linearity would otherwise be demanded by gauge-invariance under an abelian Lie algebra, such as that trivially generated by $U(1)$. Extrinsic curvature and elastic anisotropy are both shown to be examples of this effect.

\begin{figure*}[t]
\centering
\includegraphics[width=\textwidth]{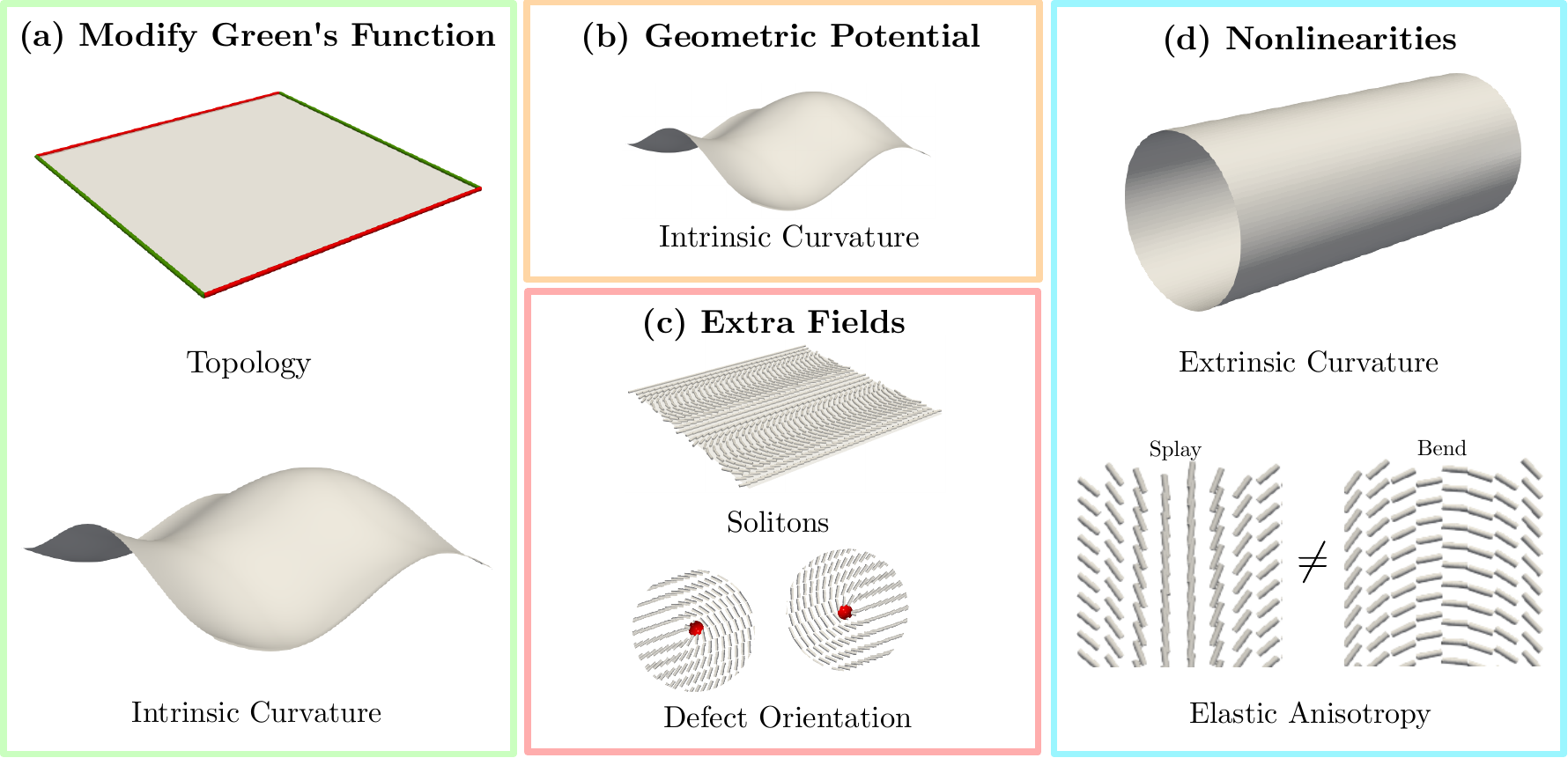}
\caption{Summary of different contributions to nematic defect dynamics on a surface. These features can (a) directly modify the interactions between defects by changing the Green's function kernel through which they interact, (b) create a potential field that defects move down the gradient of, (c) introduce an additional field with which the nematic field may interact, or (d) introduce nonlinearities to the energy, whose effect on defect dynamics can be understood perturbatively. Five of the six contributions fall neatly into one of these categories, while intrinsic curvature creates two effects, both modifing the Green's function and also introducing a `geometric potential'. }
\label{fig0}
\end{figure*}

The paper is organised as follows. We begin in Section~\ref{sec:overview} with the broad concepts. This includes a brief introduction to the ideas of abelian geometric field theory and how this relates to other approaches to defect dynamics, such as that attributed to Halperin and Mazenko \cite{liu_defectdefect_1992, mazenko_ordering_1998, mazenko_velocity_1999, angheluta_role_2021, schimmingSingularityIdentificationCharacterization2022, schimmingKinematicsDynamicsDisclination2023, schimming_defect_2025}. We also explain, at a high level, the Hodge decomposition~\cite{hodge_theory_1989}, and describe how it implies three classes of fields, as described above, as well as how it endows the energy with a topological structure. We go on to give a generic summary of perturbations to the energy that give rise to nonlinearities in the Euler--Lagrange equations. In the remainder of the paper we focus in detail on the four classes of effects shown in Fig.~\ref{fig0}. In Section~\ref{sec:dynamics}, we calculate how certain features lead to modifications to the first of the Hodge fields, and hence the propagator. These are the global topology of the surface and its intrinsic curvature. The geometric potential arising from intrinsic curvature is then discussed in Section~\ref{sec:potential}. We go on to describe, in Section~\ref{sec:extra}, the effects that give rise to the two fields in the Hodge decomposition that would not be present in a gauge invariant theory. These correspond to topological solitons and defect orientation, for each of which we provide explicit calculations of the effect on interactions. Finally, in Section~\ref{sec:nonlinear}, we describe how extrinsic curvature and elastic anisotropy modify the Euler-Lagrange equations such that non-linearities give rise to many-body effects. We conclude in Section~\ref{sec:discussion} with a summary of our results, a discussion of the benefits of our geometry-centric approach---clarity, and therefore extension to the more complicated setting of three-dimensional nematics, as well as relative ease with which the approach can be adapted for numerical studies---and brief overview of how this might be extended to incorporate flows, activity and deformable surfaces.

\section{Geometric Field Theory of Defects in 2D}
\label{sec:overview}

A nematic material is traditionally represented by its director field ${\bf n}$. This is equivalent to using an order parameter field $\Psi$, valued in a Lie group $G$ that acts on a reference direction, ${\bf n}_0$. Specifically, at each point ${\bf n}({\bf r}) = g^{-1}{\bf n}_0g$, with the group element $g=\Psi({\bf r})$ acting on the fixed reference vector in the tangent space by rotations to obtain the nematic director. 

To see this explicitly for a flat 2D plane, note that the nematic director field can generically be written as ${\bf n} = \cos \theta {\bf e}_x + \sin \theta {\bf e}_y$. If $G=U(1)$, the order parameter is then $\Psi({\bf r}) = \exp{i\theta({\bf r})}$, which implicitly sets ${\bf n}_0 = {\bf e}_x$. As we will see, on a curved surface, we will have to choose a different reference direction ${\bf n}_0$ that will not be constant. For this, we will restrict ourselves to surfaces $M$ that are either closed (a torus or sphere) or extended with a uniformity constraint imposed on ${\bf n}$ at infinity.

\subsection{Nematics as a $U(1)$ Field Theory}
In geometric field theory, the state of a material with order parameter valued in a group $G$ is represented by $\mathfrak{g}$-valued 1-form $A$ on the manifold $M$, where $\mathfrak{g}$ is the Lie algebra of $G$. Such a 1-form always defines a connection $D_A$, with associated curvature form $F = D_A A$. The relationship between $A$ and $F$ is nonlinear in the case where $G$ is non-abelian.

A standard way to construct $A$ from the order parameter field is to pull back the Maurer--Cartan form $\omega$ on the group $G$ via the map $\Psi$, defining $A = \Psi^*\omega$~\cite{hamilton_mathematical_2017}. Provided that the map $\Psi$ is smooth, this is a `pure gauge' connection with vanishing curvature $F=0$. The vanishing of curvature is an integrability condition: it ensures that the `strain field' $A$ comes from a well-defined `displacement field' $\Psi$. Conversely, if $\Psi$ is not smooth, then are a defects in the field and these result in a localised, delta-function curvature at the defect positions---hence, we may interpret $F$ as a defect density.

Concretely, the relation between the strain $A$ and the order parameter field is $A = \Psi^{-1}d\Psi$. In 2D this gives:
\begin{equation}
    \begin{aligned}
        A &= e^{-i\theta}de^{i\theta}, \\
          &= e^{-i\theta} e^{i\theta} id\theta, \\
          &= i d\theta,
    \end{aligned}
\end{equation}
where $\theta$ is the director angle. In what follows we will drop the factor of $i$ that appears here---we are effectively absorbing it into the definition of the defect charges---and write $A = d\theta$ for the connection 1-form associated to the director field. 

Because $U(1)$ is abelian, the curvature is simply $F=dA$ and there are no non-abelian effects---these only manifest in 3D. Because $A$ is exact, the curvature vanishes identically for a smooth field $\theta$, but when there are defects $F$ is nonzero and will be localised at the defect positions. 

This construction can be related to the more commonly employed Halperin--Mazenko formalism~\cite{liu_defectdefect_1992, mazenko_ordering_1998, mazenko_velocity_1999, angheluta_role_2021, schimmingSingularityIdentificationCharacterization2022, schimmingKinematicsDynamicsDisclination2023, schimming_defect_2025}. In that formalism, one still begins with the order parameter $\Psi$, and then derives the defect density $\rho$, which is the determinant of the Jacobian of the map, $\rho = |T\Psi|$. This scalar field is localised at defects, and in the 2D case it is related to $F$ by
\begin{equation}
    F = \rho \, \mu,
\end{equation}
where $\mu = dx \wedge dy$ is the area 2-form on the surface. Equivalently, we can write $\rho = \star F$, where $\star$ is the Hodge star on the surface. 

\subsection{Differences with Electrodynamics}

In electromagnetism, $A$ is the `vector potential' and $F$ is the `electromagnetic field strength'. Electromagnetism, as well as the analogous theories of the weak and strong interaction, are `gauge theories'. Loosely, this requires that the theory not depend on $A$, which can be shifted by a `phase' $A \mapsto A + d\chi$ which is nonphysical. Such shifts are called gauge transformations, and gauge theories are required to only depend on quantities left invariant under gauge transformations. This strongly constrains which terms can appear in the energy---for example, the kinetic part of the energy is usually taken to be $F \wedge \star F$, which is invariant under gauge transformations.

In elastic theories, by contrast, the connection 1-form $A$ is the strain tensor and $F$ is the defect density. In linear elasticity, the stress tensor is built out of irreducible pieces of $\star A$~\cite{machonUmbilicLinesOrientational2016}, which for a 2D nematic are vector-valued forms encoding the splay and bend~\cite{nivGeometricFrustrationCompatibility2018}. The elastic energy is then a contraction between the stress and strain---in differential forms language, $A \wedge \star A$---which is manifestly {\it not} gauge invariant. This is the origin of a great deal of interesting phenomena in elastic materials, and underlies several qualitative differences with gauge-invariant field theories such as electromagnetism. These differences are expounded upon in detail throughout the rest of this paper.

\subsection{Solving the Euler--Lagrange Equations in Terms of the Propagator}
The free energy of a nematic director field ${\bf n}$ in the one elastic constant approximation is 
\begin{equation}
    E = \frac{1}{2}\int_M |\nabla {\bf n}|^2 \mu,
\end{equation}
where we have chosen to work in units where the elastic constant is equal to 1. Assuming that the surface is flat, we can always write ${\bf n} = \cos \theta {\bf e}_x + \sin \theta {\bf e}_y$, and therefore $A=d\theta$. This implies an XY-type model, in the way alluded-to above:
\begin{equation}
    E = \frac{1}{2} \int_M A \wedge \star A.
\end{equation}
We wish to minimise this energy subject to the constraint that there are defects with charges $q_j$ at positions ${\bf r}_j$, which is encoded by the condition $dA = F$, effectively a boundary condition. Varying the energy, we find that extremal solutions satisfy
\begin{equation} \label{eq:euler--lagrange}
    \begin{aligned}
        d^* A = 0, \\
        dA = F,
    \end{aligned}
\end{equation}
where $d$ is the exterior derivative, and $d^* = (-1)^{2k+1}\star d \star$ is the exterior coderivative acting on $k$-forms.
In this form, the equations are formally analogous to the spacetime Maxwell's equations, which in geometric form are $dF = 0$ and $d^*F = J$ for $J$ the current, under the Lorenz gauge where $d^* A = 0$. 

To proceed, we have to solve two equations constraining $dA$ and $d^*A$, but a convenient choice of representation allows us to combine these into one equation. By the Hodge decomposition, $A = d\chi + d^*\psi + h$, where $\chi$ is a function, $\psi$ is a 2-form, and $h$ is a harmonic 1-form satisfying $\nabla^2 h =(dd^*+d^*d)h=0$. Since we only consider closed surfaces or those of infinite extent, $h$  meets this condition by being both closed and coclosed, $dh=d^*h=0$~\cite{hodge_theory_1989}.
To satisfy $d^*A=0$, it therefore suffices for the exact part, $d\chi$, to satisfy $d^*d\chi = \nabla^2 \chi = 0$. That is to say $\chi$ is also harmonic. Again invoking the fact that harmonic forms are closed for the surfaces $M$ that we are considering, we further see that $dA = dd^* \psi = \nabla^2 \psi$. If we prescribe $F = \sum_j q_j \delta({\bf r}-{\bf r}_j) \mu$, then we find that the field $\psi$ must satisfy 
\begin{equation} 
    (dd^*+d^*d) \psi = \sum_j q_j \delta({\bf r}-{\bf r}_j)\mu.
\end{equation}
Thus, $\psi$ can be obtained by convolving the Green's function $\mathcal{G}$ of the Laplacian with the delta-source describing the defect positions. Explicitly, this Green's function solves 
\begin{equation}
    (dd^*+d^*d) \mathcal{G}({\bf r}) = \delta({\bf r}).
\end{equation}
This then gives us the solution,
\begin{equation} \label{eq:solution}
    \psi({\bf r}) = \sum_j q_j \mathcal{G}({\bf r}-{\bf r}_j)\mu.
\end{equation}
Then $\psi$ relates to the director angle $\theta$ by $d\theta = d^*\psi$, so that $A$ is equivalently given by $A=d\theta$---the vector field whose angle is $\star \psi$ is the orthogonal complement ${\bf n}_\perp$ of ${\bf n}$, manifestly equivalent from the energetic point of view, at least in the one elastic constant approximation. 

Now we introduce polar coordinates $(\rho_j, \phi_j)$ centred on the defects. Since we will ultimately be considering curved surfaces, it is worth noting that the angular coordinate in a polar coordinate system is always related to the Green's function by $d\phi_j({\bf r})  = \star d\mathcal{G}(\rho_j)$, a fact we will make use of several times. We may also write $\rho_j = \exp(-\mathcal{G})$, which gives us the relationship $\star d\log_j = d\phi_j$ (modulo a choice of normalisation). Further, note that the (standard) notion $d\phi_j$ for these angular forms can be deceptive: as $\phi_j$ is not a well-defined single-valued function these forms are not exact, and $d(d\phi_j)$ is not zero as the notation suggests, but rather a delta function, $d(d\phi_j) = \delta({\bf r}-{\bf r}_j)\mu$. 

Using these observations, Eq.~\eqref{eq:solution} may equivalently be written as
\begin{equation}
    A = \sum_j q_j d\phi_j,
\end{equation}
a superposition of angular forms around the defects. While this is {\it a} specific solution to Eq.~\eqref{eq:euler--lagrange}---in gauge theories, this is sometimes called the Lorenz gauge, as it makes the field divergence-free---it is not {\it the} solution, as we will discuss presently. For now, we derive the energy of this configuration. To do this, we first introduce the inner product on pairs of differential $p$-forms~\cite{bott_differential_1982, hodge_theory_1989}, 
\begin{equation} \label{eq:inner_product}
    \langle \alpha, \beta \rangle = \int_M\alpha \wedge \star \beta.
\end{equation}
The energy is just $ \langle A, A\rangle/2$. The exterior derivative $d$ and coderivative $d^*$ are adjoint with respect to this inner product, 
\begin{equation}
    \langle \alpha, d\beta \rangle = \langle d^*\alpha, \beta \rangle.
\end{equation}
Using this, we can quickly derive the energy of this particular solution,
\begin{equation} \label{eq:energy_simple}
    \begin{aligned}
        E &= \frac{1}{2}\langle A, A\rangle, \\
          &= \frac{1}{2}\langle d^*\psi, d^*\psi\rangle, \\
          &= \frac{1}{2}\langle dd^*\psi, \psi\rangle, \\
          &= \frac{1}{2}\langle F, \psi\rangle, \\
          &= \frac{1}{2}\sum_{i,j} \int_M q_i q_j  \delta({\bf r}-{\bf r}_i)  \mathcal{G}({\bf r}-{\bf r}_j) \mu, \\
          &= \frac{1}{2}\sum_{i,j} q_i q_j \mathcal{G}({\bf r}_i - {\bf r}_j). 
    \end{aligned}
\end{equation}
To get from the third to the fourth line, we have recognised that $F = dd^*\psi$. We then insert the known forms for $F, \psi$ into the equation to obtain the final result. Here, we are abusing notional slightly to write ${\bf r}_i - {\bf r}_j$ for the geodesic vector between the defects $i, j$, which is only a straight line on a flat plane. The terms with $i \neq j$ are finite, and give the interaction energies. The terms with $i=j$ are the self-energies of the defects, which are infinite unless we regularise---we neglect these, as our goal here is to understand only the interactions. 

The formula in Eq.~\eqref{eq:energy_simple} implies the defects interact through the Green's function. The force acting on the defect at position ${\bf r}_i$ is obtained from this by differentiating with respect to position, 
\begin{equation}
    {\bf f}_i =  -\frac{1}{2}q_i \sum_{j \neq i }q_j \nabla_{{\bf r}_i}\mathcal{G}({\bf r}_i - {\bf r}_j).
\end{equation}
On a flat 2D surface this is just the usual, and long established, Coulomb force between defects. As we will show, one aspect of generalising this result will involve modifying the Green's function to take account of \textit{e.g.}, the global topology and/or the intrinsic curvature of $M$.


\subsection{Extra Fields: Hodge Theory and the Topology of Solutions} \label{sec:hodge}
In a gauge theory like electromagnetism, acting on $A$ by a position-dependent group element $g$ produces a different connection which must represent the same physical state, and thus we are free to always work in the Lorenz gauge. In an elastic system this is not the case, as locally-varying rotations produce a physically distinct state with different energy, for example. In order that this different state still satisfy the Euler--Lagrange equations~\eqref{eq:euler--lagrange}, it follows it must correspond to shifting $A$ by a harmonic 1-form. Broadly, the space of harmonic forms reflects our freedom to `excite' the field by imposing additional constraints, such as fixing the relative orientation of defects. The modified field is then a \emph{local} minimiser of the energy under this constraint. Hodge theory~\cite{hodge_theory_1989} describes the structure of harmonic forms---\textit{i.e.}, excitations---and their relationship with the topology of the domain; our goal is to connect them to observable, physical features of the field.

We first use Hodge theory to obtain an appropriate basis for the space of harmonic forms. In the presence of defects, the material domain is not really the full surface $M$, but rather the punctured surface $M-\{ {\bf r}_1, \dots, {\bf r}_n \}$, where ${\bf r}_j$ are the defect points. Even if $M$ is topologically trivial, the punctured surface is not. In particular, writing $\rho_j, \phi_j$ for the polar coordinates around the defects, the angular 1-forms $d\phi_j$ are closed  and coclosed (in the defect complement) but not exact, and generate nontrivial cohomology classes in the punctured surface. The 1-forms $d\log\rho_j = \star d\phi_j$ are also harmonic, but topologically trivial (exact). Indeed, exact forms are always topologically trivial in the complement of the defect set, and hence this contribution can always be removed by a homotopy of the director field, and {\it will} be removed as doing so lowers the energy. Thus, they represent excitations which will coarsen away under gradient flow, not fundamental `particle-like' objects such as defects, which are topologically protected. Nonetheless, they can play a role in the dynamics of nematics, as we can use them to describe local changes in defect orientation. 

There is a further subtlety: on a {\it closed} surface $M$ of nonzero genus (e.g., a torus, but not a sphere) the nontrivial first cohomology $H^1(M, \mathbb{Z})$ implies the existence of topologically nontrivial harmonic forms even in the absence of defects, which cannot simply be removed by homotopies and represent nonsingular but still nontrivial topological excitations: these are generally known as topological solitons~\cite{manton_topological_2010}. On an orientable 2D surface $M$, we have an equivalence between homology $H_1$ and cohomology $H^1$, and also the (abelian) fundamental group: $H^1(M, \mathbb{Z}) \cong H_1(M, \mathbb{Z}) \cong \pi_1(M)$~\cite{bott_differential_1982}. This allows us to identify solitons either with global harmonic forms (which generate the cohomology), or as localised line-like features which are homotopically nontrivial on $M$ (which generate the homology and fundamental group). 

\begin{figure*}[t]
\centering
\includegraphics[width=\textwidth]{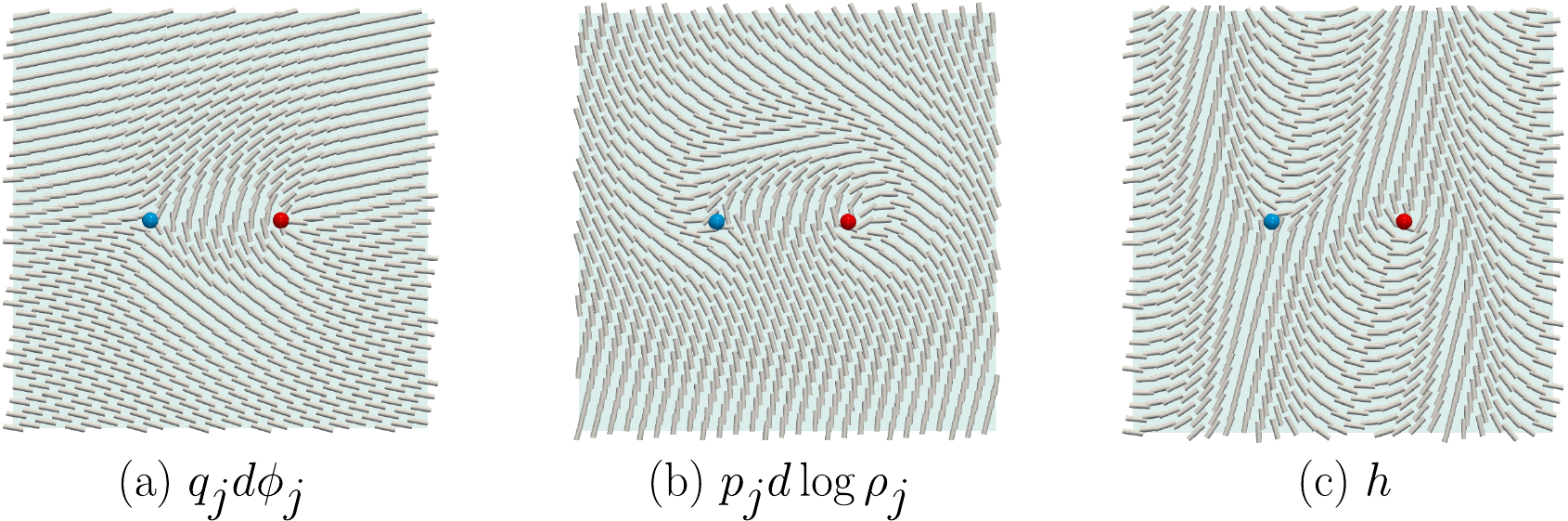}
\caption{The Hodge decomposition elucidates the different degrees of freedom in a nematic field on a torus. (a) The singular forms $q_j d\phi_j$ encode the presence of topological defects with charges $q_j$ ($q_1=+1/2$ in red, $q_2=-1/2$ in blue). (b) The additional form $p_j d\log\rho_j$ is topologically trivial, but manifests geometrically as a `spiral charge' ($p_1=p_2 = \pi/4$). (c) Nonsingular harmonic 1-forms belonging to nontrivial cohomology classes on the torus also adjust the field. Here we take $h = dx$, for $x$ one coordinate on the torus. }
\label{fig1}
\end{figure*}

Specifically, for a genus $g$ surface $M$ with $n$ defects, we have $H^1(M, \mathbb{Z}) \sim \mathbb{Z}^{2g+n-1}$. By Hodge theory, the cohomology can be generated by harmonic forms, and it follows that $2g+n-1$ then counts the topologically distinct harmonic fields, which provides a clean decomposition of the different terms which may enter into $A$. The generators are given by $d\phi_1, \dots, d\phi_n$, plus a set of harmonic generators $h_{1}, \dots, h_{2g}$ corresponding to fields that wind around nontrivial cycles on the punctured surface (note this overcounts by one). We also have a zero class of topological trivial fields, and it is convenient to represent this by a sum over the 1-forms $d\log \rho_1, \dots,d\log \rho_n $. Physically, the 1-forms $d\phi_j$ represent defects, $h_j$ represent solitons, and $d\log \rho_j$ represents orientation---the first two are topologically protected because they correspond to nonzero cohomology classes, while the latter is not. We show a pictorial representation of this decomposition in Fig.~\ref{fig1}. 

A general harmonic solution $A$ can always be decomposed in the basis we have identified, and can therefore be written in full generality as 
\begin{equation} \label{eq:general_solution}
    A = \sum_{j=1}^n q_j d\phi_j + \sum_{j=1}^n  p_j d\log\rho_j + \sum_{i=1}^{2g} h_i.
\end{equation}
Here, the $h_i$ are smooth and form a basis for the cohomology $H^1(M, \mathbb{R})$, and the index $i$ runs over the cohomology classes of the surface $M$, while the index $j$ runs over the set of defects. The field is then neatly made up of three contributions, corresponding physically to defects, orientation, and solitons. We visualise the different components of the field in Fig.~\ref{fig1} for a situation where we have a pair of defects of opposite winding on a torus. In panel (a), we show only the $(d\phi_1-d\phi_2)/2$ part of the field, with extraneous degrees of freedom set to zero. In panel (b), we introduce a extra degree of freedom $\pi(d\log \rho_1 + d\log\rho_2)/4$, with $h=0$. In panel (c), we set the orientation degree of freedom to zero but include a soliton, $h = dx$. 

Under the decomposition \eqref{eq:general_solution}, the energy is then 
\begin{equation} \label{eq:energy_solitons}
    \begin{aligned}
        E &= \frac{1}{2}\int_M A\wedge \star A, \\
          &=\text{self-energies} + \frac{1}{2}\sum_{i=1}^{2g}\sum_{j=1}^n q_j \int_M d\phi_j  \wedge \star h_i + \frac{1}{2}\sum_{i=1}^{2g}\sum_{j=1}^n p_j \int_M d\log\rho_j\wedge \star h_i + \frac{1}{2}\sum_{i,j=1}^{2g}\int_M h_i \wedge \star h_j +\\
          & \ \ \ \ \ \ \ + \frac{1}{2}\sum_{j \neq k} \int_M q_j q_k d\phi_j \wedge \star d\phi_k + p_jp_k d\log \rho_j \wedge \star d\log \rho_k + q_jp_kd\phi_j \wedge \star d\log \rho_k. \\
    \end{aligned}
\end{equation}
Each term is nothing more than the inner product on differential forms, Eq.~\eqref{eq:inner_product}. On manifold without boundary, forms that belong to different cohomology classes are orthogonal with respect to this inner product~\cite{bott_differential_1982, hodge_theory_1989}, a fact which is easy to see from Stoke's theorem and integration by parts. This gives a generic criterion for the vanishing of part of the energy. When there is a boundary, the inner product of forms that belong to different cohomology classes reduces to a boundary integral---since in our case the boundary consists of small disks removed around the defect points, we see this as the energy localising on defects. This is the Hodge theoretic perspective on why our system can be reduced to a `coulomb gas' of interacting particles. 

In what follows we will neglect the self-energies of the defects and solitons, and consider just the interaction part of the energy. The soliton self-energy is finite and easily calculated, while the defect self-energy is infinite and must be properly regularised. We can then evaluate each of the interaction integrals straightforwardly using Stokes' theorem. The interaction between the harmonic forms $h$ and the defects is 
\begin{equation}
    \begin{aligned}
        \int_M h_i \wedge \star d\phi_j &= \int_M d\phi_j \wedge \star h_i, \\
                                        &= \int_M d(\phi_j \star h_i), \\
                                        &= \sum_k \int_{\partial D_k} \phi_j \star h_i,\\
                                        &= 0
    \end{aligned}
\end{equation}
Here, we have used the symmetry of the inner product in the first line, integration by parts plus the harmonicity of $h_i$ in the second line, and then Stoke's theorem in the third to transform the integral into an integral over the boundaries of small disks $D_k$ about the defects. This then vanishes, because each of these boundary curves is contractible in $M$ and therefore the integral of the harmonic form over it vanishes. Using the same logic, along with the fact that $\star d\log \rho_j = -d\phi_j$, we find that the $d\rho_j\wedge \star h_i$ term also vanishes. We can also see the vanishing of this integral argument using topological arguments: by relative Hodge theory on the punctured surface, the cohomology classes $[d\phi_j]$ and $[h_i]$ are orthogonal, and hence this integral vanishes~\cite{bott_differential_1982}.

The integrals of $\phi_j \wedge \star d\phi_k$ and $d\log \rho_j \wedge \star d\log \rho_k$ are equivalent, because of the identity $\star d\log \rho_j = d\phi_j$. They give the standard coulomb interaction we have already discussed. The final integrand is 
\begin{equation}
    \begin{aligned}
        \int_M d\phi_j \wedge \star d\log \rho_k &=  \int_M d\phi_j \wedge d\phi_k , \\
                                            &=  \int_M d(\phi_jd\phi_k) - \int_M\phi_j d^2\phi_k,\\
                                            &= 2\pi \phi_j({\bf r}_k), \\
                                            &= 2\pi \phi_{kj}.
    \end{aligned}
\end{equation}
Here, we have again used Stoke's theorem, and we have defined $\phi_{kj} = \phi_j({\bf r}_k)$, a notation we will use throughout. This angle is ambiguous, because it depends on a set of branch cuts we use to define the multi-valued function $\phi_j$---we will explain how to properly understand this in Section~\ref{sec:orientation} below. 

Finally, we consider the soliton interaction term, $h_i \wedge\star h_j$. Since the $h$s are smooth harmonic forms, Hodge theory ensures this integral vanishes whenever $i \neq j$, because the cohomology classes corresponding to these forms are orthogonal. Thus, solitons contribute only self-energies. 

Combining the above observations, the total energy of a general harmonic field $A$ representing a director field becomes 
\begin{equation} \label{eq:general_energy}
    E = \sum_{j\neq k} (q_jq_k+p_jp_k)\mathcal{G}({\bf r}_j -{\bf r}_k) + 2\pi q_jp_k\phi_{kj} +\text{self-energies}, 
\end{equation}
where $\mathcal{G}$ is the Green's function. The smooth harmonic part does not modify the interaction, it simply contributes a self-energy. The interaction on a closed manifold $M$ is different from that on a flat plane purely because the Green's function is different. 

We emphasise here that we are only ever considering a state in which the field $A$ is assumed to minimise energy subject to defects at specific positions with specific orientations. If we do not assume that this is the case, there are additional contributions to the energy. One scenario in which this may occur is if we consider $M$ to be a manifold with boundary: there are certain boundary conditions we can impose that make it impossible to find a configuration that actually solves the Euler--Lagrange equations, i.e. to find a coclosed $A$ that satisfies the boundary conditions. This is an example of geometric frustration, where a true energy minimiser cannot be reached because of the geometry of the domain. Similarly, we can begin with a defect-free configuration which does not minimise the free energy, and is in a topological class (perhaps due to boundary conditions) which contains no harmonic representative, and hence cannot achieve a globally energy-minimising state without passing through an intermediate state with defects. 

We can model these situations by choosing a smooth field $h$ which is closed (so it has no curvature) but not coclosed (so it doesn't satisfy the Euler--Lagrange equations). In this case, where $h$ is not harmonic, the total energy of a state of the form $A = h + \sum_{j=1}^n q_j d\phi_j + p_j d\log\rho_j$ will contain nontrivial cross terms proportional to $\int \phi_j d\star h$ and $\int \rho_j d\star h$. These are bulk integrals, not boundary terms. This scenario will come up in situations where interesting textures are stabilised by boundary conditions, e.g. for dowser textures on an annulus~\cite{emersic_sculpting_2019, copar_microfluidic_2020}---the dowser structure is not represented by a harmonic field, only a closed one. While we will not describe this in further detail here, and continue to focus purely on energy-minimising states, we do note that a modification of our approach along these lines would be well-suited to studying transition pathways between stable states in nematic materials, a problem of some interest.

\subsection{Nonlinearities} \label{sec:perturbations}
In gauge theories, the extra terms that can be introduced to a gauge invariant (Yang--Mills) energy are highly constrained by the requirement of gauge invariance. By contrast, in our system the only requirement is that the energy remain invariant under ${\bf n} \mapsto -{\bf n}$, which allows for the introduction of a much larger number of extra terms. However, these terms introduce nonlinearities into the Euler--Lagrange equations, making them harder to solve.

The most natural way to introduce nonlinearity is by elastic anisotropy, assigning a different elastic constant to each invariant piece of the gradient tensor~\cite{machonUmbilicLinesOrientational2016, pollardIntrinsicGeometryDirector2021}, the result being the most general form of the Frank energy~\cite{degennesPhysicsLiquidCrystals2013}. As we will see, accounting for extrinsic curvature also results in a nonlinear coupling between the director angle and the curvature anisotropy. 

Because these examples share a common structure, it is worth treating the general case here, and specialising to the specific cases later. We are then motivated to consider an energy of the general form
\begin{equation}
    E = \frac{1}{2}\int_M \left( |\nabla \theta|^2 +\epsilon f(\theta, \nabla \theta) \right)\mu, 
\end{equation}
where $\epsilon$ is a parameter which we will usually understand to result from some form of anisotropy, and $\mu$ again denotes the area form on $M$. This structure occurs when we have distinct Frank constants (elastic anisotropy) and also when we account for extrinsic curvature (curvature anisotropy). One might also consider higher-order terms in the energy, as has been suggested for some classes of nematic material~\cite{paparini_elastic_2024, barbero_fourthorder_2019}, but we do not do so here. 

We will still seek solutions of the form $A =d \theta$, but the Euler--Lagrange equations are now nonlinear and cannot be solved directly using a Green's function. Instead, we are forced to result to perturbation theory, assuming that $\epsilon$ is small. We perform an expansion $\theta = \theta_0 +\epsilon \theta_1 +O(\epsilon^2)$. The full Euler--Lagrange equation is
\begin{equation}
    \nabla^2 \theta + \frac{\epsilon}{2}\left(\nabla \cdot \frac{\partial f}{\partial \nabla \theta}- \frac{\partial f}{\partial \theta} \right)= 0
\end{equation}
Now we insert the perturbative expansion and look at the constant and $O(\epsilon)$ parts. From this, we find that the Euler--Lagrange equations for the first two contributions are 
\begin{equation} \label{eq:nonlinear_EL}
    \begin{aligned}
        \nabla^2 \theta_0 = \sum_j q_j \delta({\bf r}-{\bf r}_j), \\
        \nabla^2 \theta_1 = \frac{1}{2}\left(\frac{\partial f}{\partial \theta} - \nabla \cdot \frac{\partial f}{\partial \nabla \theta}\right)(\theta_0, \nabla \theta_0)
    \end{aligned}
\end{equation}
We see that the linear order correction $\theta_1$ is obtained by convolving the Green's function for the Laplacian with a source that depends only on $\theta_0$. The pattern for higher-order corrections is the same, should we wish to calculate these. The energy decomposes as 
\begin{equation}
    E = \frac{1}{2}\int_M |\nabla \theta_0|^2\mu + \epsilon \int_M \left(\nabla \theta_0 \cdot \nabla \theta_1 \right) \mu+ \frac{\epsilon}{2}\int_M f(\theta_0, \nabla \theta_0) \mu+ O(\epsilon^2).
\end{equation}
 We can use integration by parts repeatedly to evaluate the second term. Firstly, we find that
\begin{equation}
    \int_M \left(\nabla \theta_0 \cdot \nabla \theta_1\right) \mu = -\int_M \left(\theta_0 \nabla^2 \theta_1\right) \mu + \int_{\partial M} \left(\theta_0 {\bf N}\cdot \nabla \theta_1 \right)ds,
\end{equation}
where ${\bf N}$ is the normal to small disks around the defects (assuming the domain $M$ has no boundary apart from the disks we have removed around defects) and $ds$ is arclength around the boundary. Now, we use the Euler--Lagrange equations for $\theta_1$ to obtain 
\begin{equation}    
\begin{aligned}
    \int_M \left(\theta_0 \nabla^2 \theta_1\right)\mu &= \frac{1}{2}\int_M \theta_0 \left(\frac{\partial f}{\partial \theta} - \nabla \cdot \frac{\partial f}{\partial \nabla \theta}\right)(\theta_0, \nabla \theta_0)\mu, \\
        &= \frac{1}{2}\int_M \theta_0 \frac{\partial f}{\partial \theta}(\theta_0, \nabla\theta_0)\mu - \frac{1}{2}\int_M \theta_0\nabla \cdot \frac{\partial f}{\partial \nabla \theta}(\theta_0, \nabla\theta_0)\mu, \\
        &= \frac{1}{2}\int_M \theta_0 \frac{\partial f}{\partial \theta}(\theta_0, \nabla\theta_0)\mu + \frac{1}{2}\int_M \nabla \theta_0 \cdot \frac{\partial f}{\partial \nabla \theta}(\theta_0, \nabla\theta_0)\mu - \frac{1}{2}\int_{\partial M} \theta_0 {\bf N}\cdot \nabla  \frac{\partial f}{\partial \nabla \theta}(\theta_0, \nabla\theta_0)ds
\end{aligned}
\end{equation}
Therefore, we find that to linear order in $\epsilon$ the energy is 
\begin{equation} \label{eq:energy_expansion}
    \begin{aligned}
    E &= \frac{1}{2}\int_M |\nabla \theta_0|^2\mu - \frac{\epsilon}{2} \left( \int_M \theta_0 \frac{\partial f}{\partial \theta}(\theta_0, \nabla\theta_0)\mu + \int_M \nabla \theta_0 \cdot \frac{\partial f}{\partial \nabla \theta}(\theta_0, \nabla\theta_0)\mu\right) + \frac{\epsilon}{2}\int_M f(\theta_0, \nabla \theta_0)\mu \\
    & \ \ \ \ \ \ \ \ + \frac{\epsilon}{2}\sum_k \int_{\partial D_k}\theta_0 \star d\theta_1  + \frac{\epsilon}{2}\sum_k \int_{\partial D_k}\theta_0 \star d \frac{\partial f}{\partial \nabla \theta}(\theta_0, \nabla\theta_0)+ O(\epsilon^2),
    \end{aligned}
\end{equation}
for $\theta_0$ the $\epsilon=0$ solution, and $D_k$ small disks about the defect points. 

The only troublesome part is the boundary integral, which involves $\theta_1$ directly---all other parts are expressed purely in terms of the $\epsilon=0$ solution. Provided that $\theta_1$ is smooth and well-behaved at the defect points themselves, this boundary term will vanish in the limit that the disk sizes go to zero. This will be the case when the source term is smooth, for example. 

Additional simplifications occur when $f$ is homogeneous in its arguments, i.e. when $\theta \partial_\theta f(\theta, \nabla \theta) = kf(\theta, \nabla \theta)$ for some integer $k$, and likewise for $\nabla \theta$. This way, the integrals can be reduced to just integrals of $f(\theta_0, \nabla \theta_0)$. This will be the case for elastic anisotropy, as we will see presently, but not for curvature anisotropy. 

A side effect of nonlinearity is that it will introduce many-body interactions. The energy $|\nabla \theta_0|^2$ is purely quadratic in its argument, so while we have cross terms that involves pairs of charges $q_i q_j$, we will never have cubic terms. This is not the case for a general function $f$---for example, $f(\theta) = \cos(\theta)$ applied to $\theta_0 = \sum_j q_j \phi_j$ will yield terms involving all charges. Nonlinearities arise in isotropic elastic field theories associated with non-abelian groups, because in such theories the differential operator appearing in the Euler--Lagrange equation is nonlinear, but not in a $U(1)$ theory. In our case, the nonlinearities arise due to anisotropy.

\section{Topology and Geometry Alter the Propagator}
\label{sec:dynamics}
\subsection{The Flat, Infinite Plane}
On a flat, infinite plane the defect-defect interactions are obtained by substituting the Green's function for the Laplacian into Eq.~\eqref{eq:energy_simple}, which yields
\begin{equation} \label{eq:2d_infinite_flat}
    E = \frac{1}{2\pi}\sum_{i,j} q_i q_j \log |{\bf r}_i - {\bf r}_j|,
\end{equation}
the usual and familiar Coulomb interaction. Differentiating with respect to defect position then gives the familiar expression for the force between two optimally aligned nematic defects on a flat, infinite plane. 


\subsection{Topologically Non-trivial Surfaces}
The correct form for defect-defect interactions on surfaces with nontrivial topology, which includes non-closed surfaces such as an infinite cylinder, but also closed surfaces such as spheres and torii, is also derived from Eq.~\eqref{eq:energy_simple} by substituting in the appropriate Green's function for the Laplacian on the surface. Curvature of the surface modifies this interaction, as we will describe in more detail presently, but the interaction will still generally be different even when the surface metric is flat, e.g. for a flat torus, as the surface's topology effects the Laplacian and its spectrum. Locally, for very close defects, the surface Laplacian reduces to the usual 2D logarithmic interaction, but for defects at larger separation there is a correction. 

\begin{figure*}[t]
\centering
\includegraphics[width=\textwidth]{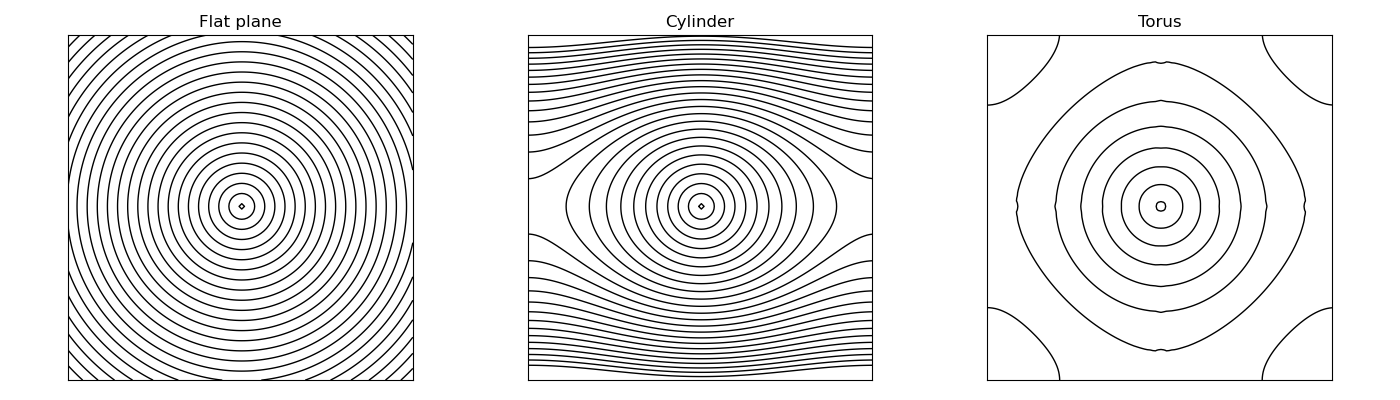}
\caption{Contour lines $\mathcal{G} = \text{const}$ for the Green's function of the Laplacian on a flat plane, a cylinder (periodic along x) and a torus (periodic along x and y) on a domain of size $2\pi \times 2\pi$, with the origin at the centre. The contour values are chosen so that they are equally-spaced curves for the flat space Green's function, and then the same values are used in each case.}
\label{fig2}
\end{figure*}

A domain of interest is a flat torus, i.e., a domain with periodic boundaries, typically employed for numerical simulations~\cite{schimming_defect_2025}. Here, curvature plays no role, but the fact that the surface is closed still modifies the Green's function to 
\begin{equation} \label{eq:gf_torus}
    G_\text{per}({\bf r}-{\bf r}^\prime) = \frac{1}{2\pi}\sum_{(m,n) \neq (0,0)} \frac{\cos(2\pi (m,n)\cdot ({\bf r}-{\bf r}^\prime))}{m^2+n^2} . 
\end{equation}
We may also consider a cylinder of radius $R$ with a standard embedding inside $\mathbb{R}^3$, e.g., with an axis along the $x$-direction and cross-section a circle in the $yz$-plane. This is an example of a surface which is not closed but still has nontrivial topology, and like the flat torus it is also a surface which has vanishing Gaussian curvature. Denoting the coordinates on the cylinder by $(x,\phi)$, the Green's function is, 
\begin{equation}
    \mathcal{G}_\text{cyl}(x-x^\prime, \phi-\phi^\prime) = \frac{1}{4\pi}\log\left( \cosh\frac{x-x^\prime}{R} -\cos (\phi-\phi^\prime)\right).
\end{equation}
For comparison, we show contour lines $\mathcal{G} = \text{const}$ for the Green's functions for the flat plane, cylinder, and torus in Fig.~\ref{fig2}. In each case we show a domain of size $2\pi \times 2\pi$, which is periodic along $x$ for the cylinder and along both $x$ and $y$ for the torus. 

We also consider, as a further example, the sphere of radius $R$. This has Green's function,
\begin{equation}
    G_\text{sph}({\bf r}-{\bf r}^\prime) = \frac{R^2}{2\pi} \log \left( \sin \frac{d_g({\bf r}, {\bf r}^\prime)}{2}\right)
\end{equation}
where $d_g({\bf r}, {\bf r}^\prime) = \cos^{-1}({\bf r}\cdot {\bf r}^\prime)$ is the geodesic distance between the points ${\bf r}, {\bf r}^\prime$ on the sphere. This then accounts for both the fact that the surface is closed and not infinite in extent, as well as the curvature, which appears because we use the geodesic distance rather than the Euclidean distance between points on the sphere. In this case however, the energy will be pick up additional terms resulting from the curvature which are not present for the flat torus and cylinder, and it is these we turn to now.

\subsection{Intrinsic Geometry} \label{sec:intrinsic}
The effect of intrinsic curvature on interactions between vortices in the XY model has been well-studied~\cite{vitelli_anomalous_2004, vitelli_defect_2004, turner_vortices_2010}. While these works consider only polar materials, we should note that 
the extension to general p-atics is largely straightforward---the only thing that changes is that generic defects have charge $\pm 1/p$, and hence the interactions are scaled by a factor of $p^2$ as opposed to the $p=1$ case. 

In the presence of intrinsic geometry, we must account for changes to the local frame field. In the standard setting, this is codified by the use of the covariant derivative in the elastic strain, $\nabla {\bf n}$.
Here, we use the connection 1-form, $\omega$. This is attractive because, like the geometric field $A$, $\omega$ is an $\mathbb{R}$-valued 1-form, and thus we can interpret it physically as being a fixed `background field' that has no dynamics of its own, but nonetheless interacts with the field $d\theta$ describing the nematic. In this way, it can formally be treated in the same fashion as one would treat a background magnetic field, with the nematic defects behaving analogously to charged particles that are no longer `free', but moving in the presence of that field. 

It is convenient to describe $\omega$ using Cartan's method of moving frames. Choose an orthonormal reference frame ${\bf e}_1, {\bf e}_2$ on the surface, and write the director field as ${\bf n} = \cos \theta \, {\bf e}_1 + \sin \theta {\bf e}_2$. Write $e^1, e^2$ for the dual coframe. Then $\omega$ is~\cite{nivGeometricFrustrationCompatibility2018, pollardIntrinsicGeometryDirector2021, dasilvaMovingFramesCompatibility2021} 
\begin{equation}
    \omega = be^1 + se^2,
\end{equation}
where $s = \nabla \cdot {\bf e}_1$ is the splay and $b = |{\bf e}_1 \cdot \nabla {\bf e}_1|$ the curvature of the ${\bf e}_1$ direction. In this frame, we see that
\begin{equation}
    |\nabla {\bf n}|^2 = |d\theta + \omega|^2. 
\end{equation}
Thus $A = d\theta + \omega$. The curvature $F$ of the field is likewise modified to $F = dA = d(d\theta) + d\omega$. If there were no defects, we would have $F = d\omega = K\mu$, where $K$ is the Gaussian curvature and $\mu$ the area 2-form. To get the appropriate expression for the defects, we have to correct for this in our boundary value for the curvature, so that our desired curvature field (boundary condition) is,
\begin{equation}
    F = \sum_j \left[ q_j\delta({\bf r}-{\bf r}_j)  -K\right] \mu.
    \end{equation}
Applying the same ideas as in the flat case, we perform the Hodge decomposition $A = d\phi + d^*\psi + h$, and then we find that 
\begin{equation}
    \nabla^2_\text{LB} \psi =  \sum_j \left[q_j \delta({\bf r}-{\bf r}_j) - K \right]\mu 
\end{equation}
where $\nabla^2_\text{LB}$ denotes the Laplace--Beltrami operator, the metric Laplacian on the curved surface. The solution $\theta$ then involves convolving with the Green's function $\mathcal{G}_\text{LB}({\bf r})$ of this operator.
In terms of geodesic polar coordinates $(\rho_j, \phi_j)$ about the point ${\bf r}_j$, we have $d\phi_j = \star d\mathcal{G}_\text{LB}$,
and thus the solution is as in the flat case, but now written in terms of {\it geodesic} polar coordinates rather than the flat space ones. 

To derive the Green's function $\mathcal{G}_\text{LB}$, we may use the fact that locally any 2D surface is conformally equivalent to a flat plane~\cite{carmo_differential_2016}. Write $x_1,x_2$ for the usual Euclidean coordinates on the infinite plane. On our curved surface $M$, locally there exist coordinates $(u_1(x_1,x_2),u_2(x_1,x_2))$ in which the metric tensor can be written as 
\begin{equation}
    g_{ij} = e^{2\chi(u_1,u_2)}\delta_{ij}.
\end{equation}
The curvature is contained within the conformal factor $\chi$. In these isothermal coordinates, the Green's function solves 
\begin{equation}
    \nabla^2 \mathcal{G}_\text{LB}  =\frac{1}{\sqrt{\det g}}\delta,
\end{equation}
where $\nabla^2$ is the flat-space Laplacian and $\sqrt{\det g}$ is the area. As long as we know the map $(u_1(x_1,x_2),u_2(x_1,x_2))$ under which we obtain our isothermal coordinates, we can solve this and then translate back into the usual coordinates $x_1, x_2$ via the inverse map to obtain the correct form of the Laplacian. Particular examples of interest are calculated in Refs.~\cite{vitelli_anomalous_2004, vitelli_defect_2004, turner_vortices_2010}.

To gain some general insight into how curvature modifies the Green's function, we note that we have an expansion, 
\begin{equation} \label{eq:LB_GF_approximate}
    \mathcal{G}_\text{LB}(\rho_j) \approx \frac{1}{2\pi}\log(\rho_j) - \frac{K({\bf r}_j)}{24\pi}\rho_j^2 + O(\rho_j^3),
\end{equation}
in the neighbourhood of some point ${\bf r}_j$, where $\rho_j$ denotes the radial polar coordinate about ${\bf r}_j$. This formula is obtained by performing a general Taylor expansion for the metric $g$ in geodesic polar coordinates about ${\bf r}_j$, and then using this to similarly expand the Laplacian~\cite{carmo_differential_2016}. It follows that, for defects that are not too widely separated, the correction to the usual flat-space Green's function decays like $\rho^2$ with a prefactor that is given by the value of the Gaussian curvature at the defect point. 

\begin{figure*}[t]
\centering
\includegraphics[width=\textwidth]{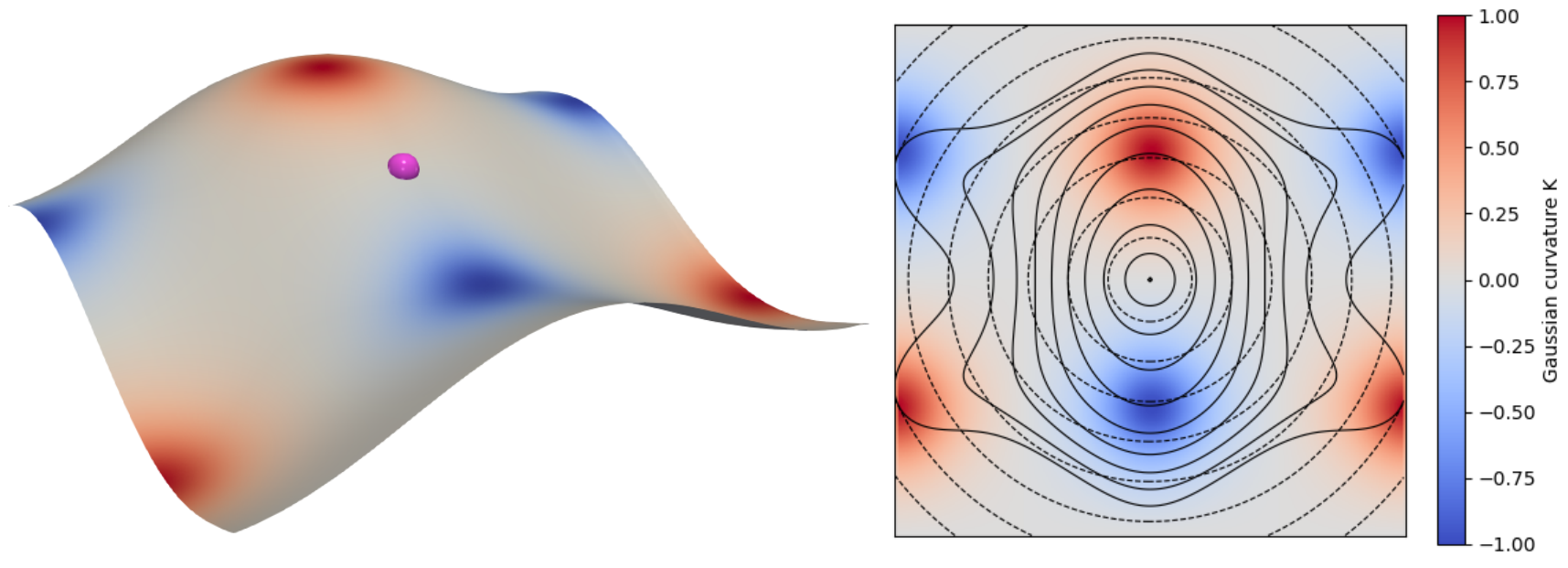}
\caption{Contour lines $\mathcal{G} = \text{const}$ for the Green's function of the Laplacian on a curved surface. We consider a surface with alternating regions of positive (red) and negative (blue) curvature. Placing a defect (magenta sphere) allows us to examine the effect of Gaussian curvature on the force felt by a second defect. The contours lines of the flat-space Green's function are shown as dashed lines. The solid lines are the contours of the curved-space Green's function, with contours corresponding to the same values as in the flat case.}
\label{fig3}
\end{figure*}

In practice, it is often convenient to express a surface in a (local) Monge gauge, in which the surface is given by the set of points $(x, y, h(x,y))$ for $h$ a height function. In Monge gauge, The surface metric $g$ is related to the flat metric $g_0$ by the relationship $g = e^{2\chi} \sqrt{\det g} g_0$, with conformal factor 
\begin{equation}
    \chi(x,y) = \frac{1}{4}\log \left(1+ (\partial_x h)^2 + (\partial_y h)^2 \right).
\end{equation}
We then deduce that 
\begin{equation}
    K = -e^{-2\chi}\nabla^2 \chi + O(|\nabla h|^4).
\end{equation}
In the limit where $\partial_x h, \partial_y h$ are small, we have the approximation
\begin{equation}
    K \approx - \frac{1}{4} \nabla^2 \left((\partial_x h)^2 + (\partial_y h)^2 \right),
\end{equation}
where we have used $\log(1+\epsilon) \approx \epsilon$ for $\epsilon$ small. To illustrate the effect of curvature, consider the surface defined by 
\begin{equation}
    h(x,y) = A\left(\cos(x) + \sin(y) \right),
\end{equation}
with $A$ chosen so that the Gaussian curvature runs between $-1$ and $+1$, and $x, y \in [0, 2\pi]$. We illustrate this surface in Fig.~\ref{fig3}. We place a defect at the point $(\pi, \pi)$ in the centre of the surface, and show the difference between the flat-space Green's function (dashed lines) and the Green's function on the curved surface (solid lines). Note that, while the surface we consider is periodic, we take it to have the topology of a plane and not a torus. 

On our curved surface, the energy is given by following the same reasoning as in Eq.~\eqref{eq:energy_simple}, 
\begin{equation} \label{eq:curvature_energy}
    \begin{aligned}
        E &= \frac{1}{2}\int_M |d^*\psi|^2 \mu, \\
          &= \frac{1}{2} \int_M \psi F, \\
          &= \frac{1}{2}\sum_{i,j} q_i q_j \mathcal{G}_\text{LB}({\bf r}_i - {\bf r}_j) +\frac{1}{2}\sum_j q_j \int_M \mathcal{G}_\text{LB}({\bf r}-{\bf r}_j)K({\bf r}) \mu({\bf r}) + \frac{1}{2}\int_{M\times M} K({\bf r}) \mathcal{G}_\text{LB}({\bf r}-{\bf r}^\prime)K({\bf r}^\prime) \mu({\bf r})\mu({\bf r}^\prime)
    \end{aligned}
\end{equation}
There are three terms in the energy. Let us focus on the first and third terms, leaving the second term for a moment. The first is the defect-defect interaction, 
\begin{equation} 
    E_\text{DD} = \frac{1}{2}\sum_{i,j} q_i q_j \mathcal{G}_\text{LB}({\bf r}_i - {\bf r}_j),
\end{equation}
which we see is exactly as in the flat case, except it now involves the Green's function on the curved surface, rather than the flat space Green's function. This is the only way that curvature alters defect-defect interactions. The final term is the curvature self-energy, which depends only on the metric and not the nematic configuration. It can be neglected for a fixed surface, where it will simply result in a constant shift to the energy independent of the director field configuration.

\section{The Geometric Potential} \label{sec:potential}
Now we return to the second term in Eq.~\eqref{eq:curvature_energy}. This involves only one defect charge, so it is not an interaction between defects. Rather, it is an interaction between the defect and the curvature, 
\begin{equation}
    E_\text{DC} = \frac{1}{2}\sum_j q_j \int_M \mathcal{G}_\text{LB}({\bf r}-{\bf r}_j)K({\bf r}) \mu({\bf r}).
\end{equation}
Following Refs.~\cite{vitelli_anomalous_2004, vitelli_defect_2004, turner_vortices_2010}, we notice that the integral in this expression is the `geometric potential' $V$ evaluated at the defect position ${\bf r}_j$, where $V$ solves the Poisson equation
\begin{equation} \label{eq:geometric_potential}
    \nabla^2_\text{LB} V = K,
\end{equation}
sourced by the Gaussian curvature. Thus we may write this in the more compact form
\begin{equation}
    E_\text{DC} = \frac{1}{2}\sum_j q_j V({\bf r}_j).
\end{equation}
According to our Taylor expansion for the metric and Laplacian, close to a point ${\bf r}_j$ we may write
\begin{equation} \label{eq:geometric_potential2}
    V = \frac{K({\bf r}_j)}{4}\rho_j^2 + O(\rho_j^3),
\end{equation}
which illustrates that this defect-curvature interaction comes in at the same order as the curvature-based correction to the defect-defect interaction term, with a curvature-dependent prefactor. Alternatively, for a surface when which is related to the Euclidean metric $g_0$ by a conformal scaling, $g = e^{2\chi} g_0$, we have $e^{2\chi} = \sqrt{\det g}$, and hence the exact identity
\begin{equation}
    K = -\frac{1}{2} \nabla^2_\text{LB}\log \sqrt{\det g}
\end{equation}
Then we see that $V$ is exactly
\begin{equation}
    V = -\frac{1}{2}\log \sqrt{ \det g} = -\chi.
\end{equation}
This convenient formula is useful for numerical simulations, where we can assume that our surface (or at least patches of our surface, in the case of a sphere) is conformally equivalent to a flat surface. Further, if we wish to extend this framework to treat deformable surfaces, the best way to do this is to treat the conformal factor as a dynamical variable, in which case the formulas we have just given allow us to describe the curvature and geometric potential quite simply. 

This further suggests an interesting analogy with 2D gravity, where fluctuations in space only manifest as conformal rescaling of the metric~\cite{difrancescoConformalFieldTheory1997}. The dynamics is given by minimising the Liouville action, a conformal scalar field theory whose dynamical variable is the conformal factor $\chi$; we see that the geometric potential is then exactly $-\chi$. Further, the energy obtained from minimising the Liouville action subject to the field having localised sources (the vertex model) is exactly $E_\text{DD} + E_\text{DC}$, with the sources being analogous to the defects in our model. The analogy is not completely exact: in 2D gravity there is only one field, $\chi$, while for us the nematic and the metric are two distinct fields which are coupled.

Alternatively, if the surface is described in Monge form with height function $h$, the geometric potential is then approximately 
\begin{equation}
    V \approx - \frac{1}{4}  \left((\partial_x h)^2 + (\partial_y h)^2 \right),
\end{equation}
again, assuming the surface is close to flat and the gradients of $h$ are small. 

The force acting on the defect at position ${\bf r}_j$ is obtained by differentiating $-E_\text{DD} - E_\text{DC}$ with respect to the position. We obtain the force, 
\begin{equation}
    {\bf f}_j = -\frac{q_j}{2} \left(\nabla_{{\bf r}_j} V({\bf r}_j) + \sum_{i\neq j} q_i\nabla_{{\bf r}_j}\mathcal{G}_\text{LB}({\bf r}_i - {\bf r}_j)\right).  
\end{equation}
The intuition here is: locally, defects interact logarithmically along geodesics; globally, they feel a `potential' $V$ and move along the gradient of this potential. In terms of an analogy with electromagnetism, this is like a `background field' which the point charges move through. Positively-charged defects will be drawn to regions of positive curvature, and negatively-charged defects will be drawn towards regions of negative curvature. 

Curvature then does the following: (i) It modifies the defect-defect interactions themselves by changing the structure of Greens' function; (ii) it contributes an additional defect-curvature interaction; (iii) it introduces a background `curvature energy', which is a constant contribution that can be neglected on a {\it fixed} surface, but becomes relevant on a deforming surface. It would be interesting to study the problem of a nematic on a deformable surface from this point of view, by introducing additional terms to the energy to account for the surface tension and bending modes of the surface itself and viewing the nematic texture in terms of the `defect gas' model we are deriving here. Potentially, the metric energy itself could be localised in terms of curvature singularities, analogous to the vertex model of 2D gravity~\cite{difrancescoConformalFieldTheory1997}. However, to do so properly means accounting for the role of extrinsic curvature, which we discuss later in Section~\ref{sec:nonlinear}.

\section{Extra Fields}\label{sec:extra}

\subsection{Solitons}
Topological solitons are smooth, defect-free fields that possess a topological invariant distinguishing them from the ground state~\cite{manton_topological_2010}. In liquid crystals, they typically take the form of localised structures: merons and Skyrmions are quasi-2D features associated to a nontrivial Euler class, while Hopfions and heliknotons correspond to knotted and linked field lines that generate a nontrivial Hopf invariant~\cite{manton_topological_2010, wuHopfionsHeliknotonsSkyrmions2022}. 

The Hodge decomposition invites us to regard solitons in 2D as smooth, global harmonic fields, not localised particle-like objects. The associated topological invariant that prevents their removal is not the curvature of the field $A$, as is the case for defects, but rather the cohomology class of the harmonic form. This can be computed via a winding number which, especially in the particle physics literature, is referred to as a `Wilson loop'~\cite{hamilton_mathematical_2017}. The Wilson loop about a closed curve $\gamma$ is the integral, 
\begin{equation}
    W[\gamma] = \int_{\gamma} A.
\end{equation}
When $\gamma$ encloses a defect, this integral computes the defect winding. When $\gamma$ is a nontrivial cycle on the torus, the Wilson loop returns an integer winding of the field along the loop, a count of solitons. 

A more topological perspective connects these numbers to the homotopy class of the director field~\cite{farber_topology_2004}. As we outlined in Section~\ref{sec:hodge}, nematic textures with $n$ defects on $M$ are classified by elements of the homology $H_1(M-\Gamma_n)$, where $\Gamma_n = \{ {\bf r}_1, \dots, {\bf r}_n \}$ denotes the set of $n$ defect points. For $M$ a genus $g$ surface, this group is isomorphic to the free group on $2g+n-1$ generators. Thus, the homotopy class of a texture in is given by the element $[w_i\gamma_i] \in H_1(M-\Gamma_n)$, where $\gamma_i$ are a set of $2g + (n-1)$ oriented curves generating the homology and the $w_i$ are the winding numbers along those curves. For a torus a natural choice consists of the usual two generators for the torus, plus $n-1$ additional parallel curves that separate the defects. For any surface $M$ the resulting group homomorphism $W : H^1(M-\Gamma_n) \to \mathbb{Z}$ (the Wilson loop, or `period map' in the mathematics literature) is determined by the collection of integers $W[\gamma_i]$, and is a complete homotopy invariant of the texture in the space of textures with $n$ defects~\cite{farber_topology_2004}---this of course boils down to the winding numbers of the defects plus the solitons.

On the flat torus, a basis for the harmonic 1-forms is given by $h_{n,m} = ndx + mdy$, where $n, m$ are integers and $x,y \in S^1$ are the coordinates on the torus. Shifting our field $A$ by these 1-forms introduces new, topologically distinct solutions with higher energy. These `excitations' cannot simply coarsen away, and result in consistently higher energies; this was recently explored by Schimming~\cite{schimming_defect_2025} in his analysis of defect interactions on the torus. We have already derived an expression for the energy when an additional harmonic field is included, Eq.~\eqref{eq:general_energy}. As this formula shows, solitons do not actually interact with defects, nor do they modify the interaction between defects: the interaction between defects on a flat torus is different from that on a flat infinite plane not because of solitons, but purely because the Green's function is different. Instead, solitons shift the total energy of the nematic configuration by their `self-energy'. On a flat torus, this is given by
\begin{equation}
    E_\text{SS} = \frac{1}{2}\int_M h_{n,m} \wedge \star h_{n,m} = \frac{1}{2}(n^2+m^2). 
\end{equation}

On curved surfaces, e.g., a torus embedded in Euclidean $\mathbb{R}^3$, the structure is the same but the specific harmonic forms are different, and this will result in different energy contributions. In terms of coordinates $u,v$ on the meridian and longitude on the torus, we can write the torus as the surface given by
\begin{equation}
    (x,y,z) = \left( (R_1 + R_2\cos u)\cos v, (R_1 + R_2\cos u)\sin v, R_2\sin u\right),
\end{equation}
where $R_1, R_2$ set the radii of the two circles comprising the torus. The meridional harmonic form is 
\begin{equation}
    h_u = \frac{1}{R_2}du,
\end{equation}
while the longitudinal one is 
\begin{equation}
    h_v = \frac{1}{(R_1 + R_2\cos u)}dv
\end{equation}
Set $h_{n,m} = nh_u + mh_v$ to be the linear superposition of these, as before. A short calculation shows the self-energy of a harmonic field is 
\begin{equation}
    E_\text{SS} = 2\pi^2R_1\left(\frac{n^2}{R_2^3} + \frac{m^2R_2}{(R_2^2-R_1^2)^{3/2}} \right).
\end{equation}
Thus, the scaling remains quadratic in the `count' of the solitons $n,m$, but there are now metric-dependent prefactors.

While these calculations show that solitons do not change the defect-defect interactions, at least in the director field picture, they can result in the creation of defects---the cost of meridional solitons on a torus is so high that it may be cheaper to nucleate defects to unwind the soliton---and in this way effect the global structure and dynamics of the texture. 

A final remark. We should be very careful about trying to apply the results of this section, which specifically uses the director field formalism for nematics, when we move to the Q-tensor framework. The latter introduces an additional variable $s$ that allows for the melting of nematic order. When there are only defects and no solitons, it well known that we will have $s$ equal to its constant equilibrium value everywhere outside some small core regions around defects. However, topological solitons are energetically costly and it is likely that the order will melt around them, introducing new effects coming from gradients in $s$ which we do not account for for in our purely director-based models. In the director field picture, solitons are global harmonic forms; in the Q-tensor formalism, the region about the soliton in which $s$ differs from its equilibrium value gives a finite-size core to the soliton, changing its behaviour. The core can induce interactions when it overlaps with the core region of a defect or another soliton. Beyond the simulations in Ref.~\cite{schimming_defect_2025}, we do not know of any detailed studies of such effects or of the core structure of solitons in 2D, and we have not examined these ourselves. Furthermore, we expect that the solitons will rapidly decay and nucleate defects, and these extra defects will exert forces on the original defects that will change their trajectories.

\subsection{Defect Orientation} \label{sec:orientation}
In general, defects need not be in perfect relative orientation to one another, as illustrated in Fig.~\ref{fig1}(a). They can instead be misaligned, as in Fig.~\ref{fig1}(b). The assumption of `perfect alignment' is really the neglecting of harmonic modes in the solution, $A$, to the Euler--Lagrange equations, which do not generate curvature. In a gauge theory like electromagnetism, adding such terms is a gauge freedom that does not change the energy (which depends on curvature $F$ and not the field $A$). In an elastic material, shifting the geometric field $A$ by a harmonic term \emph{does} change the energy, which is why orientation matters in a nematic but not in electromagnetism. Unlike solitons, these extra fields are topologically trivial and can be `unwound' without creating defects, but this process alters the interaction between defects in the process. 

In this section we discuss this in terms of our `extra fields' perspective, and relate this to the concept of a defect orientation vector. We also discuss the technical points that arise when trying to describe orientation on a curved surface, where naive application of results in the literature that implicitly assume flatness may lead one astray. 

\subsubsection{Orientation as a Boundary Condition}

There are several ways to incorporate orientation. Vromans \& Giomi~\cite{vromansOrientationalPropertiesNematic2016} and Tang \& Selinger~\cite{tangOrientationTopologicalDefects2017} have approached it in similar ways. A disk around the defect---interpreted as the defect core region---is removed, and a boundary condition is imposed to encode the orientational properties. The director angle $\theta$ still satisfies a Laplace equation with the defects acting as a source term, but is now subject to the boundary condition. The procedure of computing the interaction energy is the same: the energy is still given by Eq.~\eqref{eq:energy_simple}, but we use the Green' function for the boundary value problem rather than the free space Green's function.  

In simple cases, the required Green's function can again be constructed using conformal maps~\cite{davidson_conformal_2012, vafa_periodic_2024}, in a similar fashion to what is done for curvature~\cite{vitelli_anomalous_2004, vitelli_defect_2004, turner_vortices_2010}. For example, for two defects, we can conformally map the twice-punctured plane onto an annulus, and solve Laplace's equation for $\theta$ on that annulus with the given boundary conditions---harmonic functions are invariant under conformal maps in 2D, and therefore it suffices to take the Green's function on the annulus and express it in terms of the coordinates in the plane using the inverse of the conformal transformation. This is essentially the approach employed by Tang \& Selinger~\cite{tangOrientationTopologicalDefects2017}, while Vromans \& Giomi employ an image construction~\cite{vromansOrientationalPropertiesNematic2016}.

In general, the interactions between $m$ defects with arbitrary orientations are given by using the Green's function for Laplace's equation on $M-\cup_jD_j$ ($D_j$ being disks of some core radius $r_\text{core}$ centred at the defect points ${\bf r}_j$) subject to the appropriate boundary conditions of rotation by angles that set the local defect orientation. Because this Green's function will depend on the number of disks removed, and therefore on all rotation angles, changing local defect orientations smuggles many-body interactions (albeit, many-body interactions between the orientational degrees of freedom and not between the charges, which still only interact pairwise) into what is otherwise a system with only pairwise interactions. However, deriving the Green's function on a multiply-punctured plane is nontrivial. In practice we can approximate by taking each pairwise Green's function between defects to be the same as in the case of two defects, which amounts to neglecting many-body effects. We may also try to use an image construction to obtain solutions. 

\subsubsection{Orientation as a Spiral Charge Field}

An alternative strategy for describing orientation, advanced by \v{C}opar \& Kos~\cite{copar_manydefect_2024}, makes use of harmonic forms to encode the winding in terms of a `spiral charge', in a manner similar to how we have encoded solitons. This approach is far simpler to do calculations with, and fits nicely with the topological approach based on the Hodge decomposition. We adopt a variant on this approach here, and consequently our results are similar but not identical to those of Ref.~\cite{copar_manydefect_2024}. 

On a flat space, define
\begin{equation} \label{eq:spirals}
    s_j = p_jd\log(|{\bf r}-{\bf r}_j|).
\end{equation}
This is defined on the complement of the defect at ${\bf r}_j$. Following Ref.~\cite{copar_manydefect_2024} we refer to the scalar $p_j \in \mathbb{R}$ as the `spiral charge' of the defect. Unlike the defect charge this charge is not topological in nature and is not quantised. For a single defect, the spiral charge is gives exactly the roation angle of the defect on the boundary of a disk of radius $r_\text{core}$, $p_j\log (r_\text{core})$. 

If we are on a curved surface, then Eq.~\eqref{eq:spirals} is not exactly a harmonic form, and instead we must replace the logarithm with the Green's function for the Laplace--Beltrami operator on the curved surface, 
\begin{equation}
    s_j = p_j d\mathcal{G}_\text{LB}(d_g({\bf r},{\bf r}_j)),
\end{equation}
for $d_g({\bf r},{\bf r}_j)$ the geodesic distance to the defect at position ${\bf r}_j$, i.e., the geodesic radial coordinate about the defect. Locally this is given by Eq.~\eqref{eq:LB_GF_approximate}, so at small distances we recover essentially the same form as for the flat plane. 

To encode the spiral charge, we shift our gauge field to 
\begin{equation}
    A = A_0 + \sum_j p_j d\mathcal{G}_\text{LB}(d_g({\bf r},{\bf r}_j)),
\end{equation}
where $A_0$ is the solution to the problem with orientation not considered, which may include curvature effects, solitons, etc. Of course, as the $s_j$ are closed this still solves the required Euler--Lagrange equation $dA = \sum_j q_j \delta({\bf r}-{\bf r}_j)$.

This way of encoding the defect orientations then has them appear as an extra `field', much like solitons or curvature, except this field has a non-conserved geometric spiral charge rather than a conserved soliton number. As with this other case, the correction to the interaction doesn't appear in the Green's function, which is unchanged. Recalling Eq.~\eqref{eq:general_energy}, we see that the interaction energy between the defect charge and the spiral charge is 
\begin{equation} \label{eq:defect_orientation_interaction}
    E_{DO} = \sum_{i\neq j}q_i p_j \phi_{ji}.
\end{equation}
Here, $\phi_{ji} = \phi_{i}({\bf r}_j)$.

This formula is structurally analogous to the Aharanov--Bohm effect in electromagnetism: it is the phase generated by transporting the charge $p_j$ in a loop around the particle with charge $q_j$. It can also be understood as a form of anyon-like braiding~\cite{mietke_anyonic_2022}: smoothly interchanging a pair of defects results in a non-trivial $p_j$ for each defect. Neither analogy can be considered exact, as the spiral charge $p_j$ is neither quantised nor topological. 

The self-energy associated with the spiral charge field is distance-dependent and identical to the self-energy associated with the defect charges:
\begin{equation}
    E_{OO} = 2\pi\sum_{ij} p_i p_j \mathcal{G}_\text{LB}({\bf r}_i-{\bf r}_j).
\end{equation}
Unlike the defect charges, the spiral charges $p_j$ can smoothly go to zero, so this energy can be minimised both by bringing the defects together, and also by unwinding the spiral charges. 

\begin{figure*}[t]
\centering
\includegraphics[width=\textwidth]{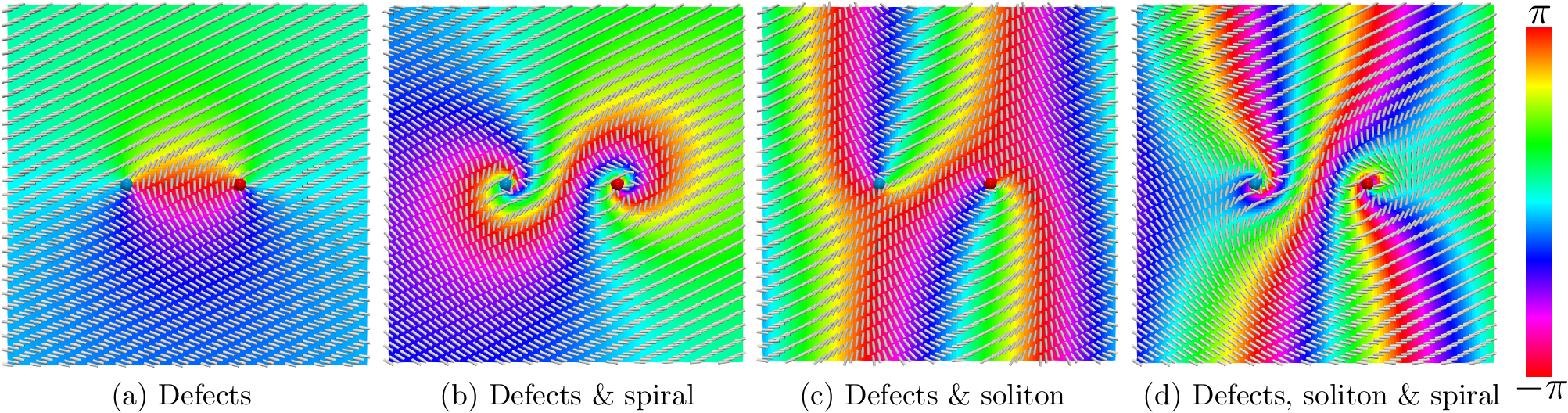}
\caption{Nematic configurations with spiral charge and solitons. In each panel we show both the director angle $\theta$ (colour) and the director itself (white sticks). (a) Two defects ($+1/2$ in red, $-1/2$ in blue) on a torus with neither spiral charge nor solitons, with field $A_0 = (d\phi_1 - d\phi_2)/2$. (b) We introduce a spiral charge to the configuration shown in panel (a), so the field is now $A = A_0 + p(d\log \rho_1 - d\log \rho_2)$. (c) We introduce a soliton to the configuration shown in panel (a), so the field is now $A = A_0 + dx$. (d) We take both the spiral charge and the soliton, $A = A_0 + p(d\log \rho_1 - d\log \rho_2) + dx$. }
\label{fig4}
\end{figure*}

In Fig.~\ref{fig4}(b) we show a pair of defects on a torus with $p_1 = -p_2$. These charges are chosen so that the defects are rotated by an angle $\pi/2$ on the boundary of a disk of radius $r_\text{core} = 0.25$. We show the colour winding $\theta$. Picking any curve $\theta = \text{const}$ gives a curve that connects the two the two defects. We can consider homotopy classes of curve whose endpoints are fixed to be the two defect sets, and from this we can see the non-topological nature of spiral charge: changing the spiral charge changes the shape of this curve but not its homotopy class, as can ready seen by examining, for example, the yellow curve in Figs.~\ref{fig4}(a,b). 

These homotopies are relevant because, if the defects are held constant and the field are allowed to change around them to minimise the energy, each curve $\theta = \text{const}$ in Figs.~\ref{fig4}(b) must deform through such a homotopy to the same curve in the energy minimiser, Figs.~\ref{fig4}(a). It is clear it will go through the `minimal' such homotopy, and indeed the curve can be regarded as an elastic curve with endpoints fixed, which will obviously smoothly deform to minimise its length. If the defects are allowed to move the same is true, except the boundaries of the elastic curve are now free. From this, we can see qualitatively what was found quantitatively by Tang \& Selinger through simulations~\cite{tangOrientationTopologicalDefects2017}: the defects will move on spiral trajectories roughly equivalent to the spiraling curve $\theta = \text{const}$ of minimal length that connects them. 

Solitons can also change the orientation of defects, even without directly adding a spiral charge term in this fashion---Fig.~\ref{fig4}(c) shows an example of a field on a torus with two defects, zero spiral charge, but with the addition of the harmonic form $h = dx$. The topological nature of this change can be seen from examining a curve $\theta = \text{const}$, for example, the yellow curve in Fig.~\ref{fig4}(c). It does not belong to the same homotopy class as the same curve in Fig.~\ref{fig4}(a,b), because it now contains two components, which both wind once around the torus (the curve connecting the defects is related to the curve with the same $\theta$ value in panel (a) by an operation called a Dehn twist~\cite{pollard_morse_2025}). We can also consider a field with both a soliton and a spiral charge, Fig.~\ref{fig4}(d). The yellow curve is again topologically distinct from its equivalent in panels (a-c): it has only a single component, and also belongs to a different homotopy class from the curve connecting the two defects in panels (a-c). 

\subsubsection{Defect Orientation Vectors}
\label{sec:orientation_consistency}
There remains a subtle point to reconcile: the angles $\phi_j$ are multi-valued functions, and to make sense of them we must define a set of branch cuts, or equivalently a set of rays emanating from the defects which we use at the zero value for these angles. Thus, the angles $\phi_{ij}$ that appear in the energy, Eq.~\eqref{eq:defect_orientation_interaction}, are ambiguous. Even if we fix this set of branch cuts once and for all---for example, by deciding to measure all angles from the x-axis---the angle $\phi_{ij}$ is still ambiguous because its value will jump by $2\pi$ whenever a defect crosses the cut line. 

A resolution to this problem is to use a physically meaningful choice of direction attached to each defect, from which angles must be measured, which we denote ${\bf p}_i$. In this way, we reconcile the `extra fields' perspective on defect orientation with the notions of a defect orientation vector. This orientation vector ${\bf p}_i$ can be defined in an invariant way in terms of the gradients of the director field, and we refer the reader to Refs.~\cite{tangOrientationTopologicalDefects2017, vromansOrientationalPropertiesNematic2016, longGeometryMechanicsDisclination2021} for a full discussion.  

Let $\theta = q\phi + p\log \rho$ be the field around a defect, with $p$ the scalar spiral charge. To define an orientation from a spiral charge, we must introduce a core cutoff radius $r_\text{core}$. We may then identify the orientation vector ${\bf p}$ with the direction where $\theta = 0$ on the boundary of this disk, i.e., the defect orientation vector is ${\bf p} = \cos \phi_* \, {\bf e}_x + \sin \phi_* \, {\bf e}_y$, where the angle $\phi_*$ is defined by
\begin{equation}
    \phi_* = -\frac{p}{q}\log r_\text{core}. 
\end{equation}
This angle depends on an intrinsic lengthscale $r_\text{core}$, i.e., it is a scale-dependent property. On a curved surface one should replace the Cartesian basis with a fixed orthonormal frame field. Equivalently, the defect orientation vector defines a physically-meaningful direction from which angles may be measured. In this manner, the angles $\phi_{ji}$ may be properly and consistently defined. 

The value $\phi_{ji}$ of the angle $\phi_j$ at the position ${\bf r}_i$ will then depend on the Green's function of the curved surface. Care should be taken on how this Green's function distorts the contours of constant geodesic distance from the point defect---see the visualisation in Fig.~\ref{fig3}---and how this will effect the value of the angle. 

A similar subtlety arises when using defect orientation vectors on a curved space. One is tempted to write quantities like ${\bf p}_i \cdot {\bf p}_j$, by which we mean the metric inner product between the orientation vectors of two defects, in order to define the angle between them. However, as written this expression is ill-defined. The vectors ${\bf p}_i$ and ${\bf p}_j$ belong to two different tangent spaces, namely, the tangent spaces at the defect positions ${\bf r}_i$ and ${\bf r}_j$. Naively, one cannot compare such vectors. In order to do, we must first parallel transport the defect vector at ${\bf r}_i$ along a geodesic so that it sits at ${\bf r}_j$ (or vice-versa), and only then may we take the inner product. On a flat space, this parallel transport is `free' and the naive value of the expression ${\bf p}_i \cdot {\bf p}_j$ is correct, but this will not be the case on a general curved surface. One should be careful of this when naively employing formulas in the literature that implicitly assume flatness. 

\begin{figure*}[t]
\centering
\includegraphics[width=0.5\textwidth]{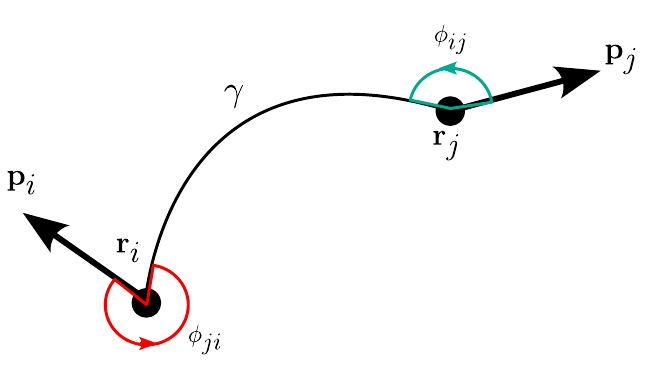}
\caption{Consistently determining the angles between defects on a curved surface. Fix two defects at positions ${\bf r}_i, {\bf r}_j$, with orientation vectors ${\bf p}_i, {\bf p}_j$. We measure angles $\phi_i, \phi_j$ anticlockwise from these orientation vectors, which are physical locations for the branch cut associated with the angles. To determine $\phi_{ij} = \phi_j({\bf r}_i)$ and $\phi_{ji}$, we must first find the geodesic $\gamma$ between them, oriented so that it is facing outward from the defect point we are considering. Then, to obtain $\phi_{ij}$ we measure the angle between ${\bf p}_j$ and the tangent vector to $\gamma$ at ${\bf r}_j$ (green arc). We do likewise for $\phi_{ji}$ (red arc).}
\label{fig5}
\end{figure*}

In terms of the spiral charge, $\phi_{ij}$ is the counter-clockwise angle, measured in the tangent space at ${\bf r}_j$, between the orientation vector ${\bf p}_j$ and the geodesic $\gamma$ from ${\bf p}_j$ to ${\bf p}_i$---see Fig.~\ref{fig5}. The way of measuring angles is simpler to implement than parallel transporting the orientation vectors, although it amounts to the same thing. We remark that `optimal' orientation, which is when $\phi_{ij} = -\phi_{ji}$, occurs on a curved surface not when the orientation vectors are colinear with one another, as on a flat space, but when they are colinear with the tangent to the geodesic between the two defect points. 

It is also worth reflecting for a moment on the pros and cons of the two approaches. One can argue that building the orientation into the Green's function is the `correct' approach, as it leaves the connection $A$ to be built entirely from topologically nontrivial parts representing topological features, defects and solitons, and also realises the orientation vectors more naturally~\cite{tangOrientationTopologicalDefects2017}. However, from the practical standpoint of doing calculations and numerical simulations, encoding the orientation in terms of extra fields is far easier in the case of multiple defects and for curved, potentially deforming, surfaces. Further, there is actually more information contained within a spiral charge field: it doesn't just account for the displacement of the defect's orientation away from a fixed axis, e.g. the $x$-axis, but how many full turns the profile of the defect makes as we move away from the defect and towards another defect, as well as the sense or handedness (clockwise or anticlockwise) of this turning~\cite{copar_manydefect_2024}. The orientation vector sees neither the number of turns nor the handedness. Thus, we may encode a broader class of fields and study their relaxation dynamics.

\section{Nonlinearities}\label{sec:nonlinear}

\subsection{Extrinsic Curvature}
To our knowledge, previous work on how curvature affects interactions between topological defects has only studied intrinsic curvature. However, work on passive nematic shells has shown that reproducing the correct behaviours requires also accounting for the extrinsic curvature of the shell~\cite{napoliExtrinsicCurvatureEffects2012}. To do this, one takes the fully 3D nematic energy, involving the splay, twist and bend, and restricts it to the surface. The energy can then be seen to decompose into an intrinsic part---equivalent to the usual 2D Frank energy on the surface and which is minimised by having the director field align with geodesics---as well as an extrinsic part, which is minimised by having the director align with the directions of principal curvature. Since defects necessarily force the director field to deviate from geodesics, it seems natural to conjecture that they will prefer to sit near umbilical points, defects in the directions of principal curvature. 

A natural way to incorporate extrinsic curvature into our description is to represent the director field as ${\bf n} = \cos \theta \, {\bf e}_1 + \sin \theta \, {\bf e}_2$ in terms of an orthonormal frame on the surface, as we did for intrinsic curvature, but then to choose the frame ${\bf e}_1, {\bf e}_2$ {\it specifically} to be aligned with directions of principal curvature. Let $\kappa_1, \kappa_2$ be the principal curvatures. We insert this ansatz for the director field into the one elastic constant 3D Frank energy. After a short calculation, we find that,
\begin{equation}
    \begin{aligned}
        |\nabla {\bf n}|^2 &= |d\theta + \omega|^2 + \kappa_1^2\sin^2\theta +\kappa_2^2\cos^2 \theta, \\
                          &=|d\theta + \omega|^2 + \frac{\kappa_1^2+\kappa_2^2}{2}+\frac{\kappa_1^2 - \kappa_2^2}{2} \cos 2\theta,
    \end{aligned}
\end{equation}
where we have used the standard trig formulae for $\cos(2\theta)$. Define the mean curvature $H = (\kappa_1+ \kappa_2)/2$, and also define the curvature anisotropy $\Delta = (\kappa^2_1 - \kappa^2_2)/2 = 2H(k_1-k_2)$. The term that is independent of $\theta$ can then be written in terms of the mean and Gaussian curvatures, and we find that 
\begin{equation} \label{eq:extrinsic_curvature_energy}
    E = \frac{K}{2} \int_M \left( |d\theta + \omega|^2 + (2H^2 - K) + \Delta \cos 2\theta \right) \mu.
\end{equation}
The first term is the intrinsic part of the energy, which we have dealt with in Sections~\ref{sec:intrinsic} and~\ref{sec:potential}, and the second and third terms represent the extrinsic contribution to the energy. The Helfrich-like term $2H^2-K$ gives a bending energy for the surface which will be important on a deforming surface, but for us is a constant contribution that we can ignore. Thus the extrinsic part of the energy reduces to a single term $\Delta \cos 2\theta$ which becomes negligible in the limit $\Delta \to 0$ of equal principal curvatures, i.e. on a sphere. This term will generally be non-negligible on, for example, a cylinder of radius $R$, where $\Delta = 2/R^2$, as well as on torii. Large cylinders and `squashed spheres', spheres deformed to be slightly elliptical, have small but nonzero $\Delta$ and therefore represent an interesting perturbative regime for analysing the effect of curvature anisotropy. 

There are specific scenarios in which this equation has trivial solutions. On a generic closed surface, the directions of principal curvature will have defects (umbilic points) which are generically of winding $\pm 1/2$. A director field with defects exactly at the principal umbilic points can then be obtained by taking a constant $\theta = \pi/4$. This constant $\theta$ seems a little confusing, as at first glance it appears to mean that there are no defects. However, defects may still occur because they appear in the frame ${\bf e}_1, {\bf e}_2$, with which the director then makes a $45^{\circ}$ angle---the energy is concentrated in $\omega$, the connection form for the surface written in this basis. The `charge' $q_j$ of the defects is only their `relative charge', or the true defect charge minus the charge of the singularity in the frame field---thus, $q_j=0$ means a defect sat at an umbilic point whose winding matches that of the umbilic. Such defects are then not expected to be motile (i.e., such a configuration represents a genuine fixed point of the free energy) in a passive system, unless the surface is deformable.

In the general case, we can find approximate solutions to Eq.~\eqref{eq:extrinsic_curvature_energy} via a perturbative approach in the limit for small $\Delta$, as outlined in Section~\ref{sec:perturbations}. In order to apply this approach, we require that $\Delta$ is constant, so that we can take it out of the integral. This will not generically be the case, although it will be the case for a cylinder, for example. For the remainder of this section, we will assume either that $\Delta$ is constant, or that its gradient is so small any variations can be neglected. 

Under this assumption, we have,
\begin{equation}
    \begin{aligned}
        f &= \cos(2\theta), \\
        \partial_\theta f &= -2\sin(2\theta), \\
        \partial_{\nabla \theta} f &= 0. 
    \end{aligned}
\end{equation}
The vanishing of $\partial_{\nabla \theta} f$ greatly simplifies the calculations we have to do. It removes one of the boundary terms in our general expansion for the energy, Eq.~\eqref{eq:energy_expansion}. The troublesome boundary term involving $\theta_1$ also vanishes, because the source $-2\sin(2\theta_0)$ ensures $\theta_1$ will be smooth at the singular points. The remaining terms in our expansion Eq.~\eqref{eq:energy_expansion} then give us,
\begin{equation} \label{eq:extrinsic_curvature_corrections}
    E = E_0 + \Delta\int_M \theta_0 \sin 2\theta_0({\bf r})d{\bf r} + \frac{\Delta}{2} \int_M \cos2\theta_0({\bf r}) d{\bf r},
\end{equation}
where $E_0$ is the energy of the standard solution on the curved surface. We take,
\begin{equation}
    \theta_0 = \sum_j q_j d\phi_j,
\end{equation}
where $\phi_j$ are understood as the angles in geodesic polar coordinates about the defects, with $d\phi_j = \star d \mathcal{G}_{LB}$ in terms of the Green's function on the surface. 

Direct evaluation of these integrals is challenging. For simulations, they can be solved numerically. Analytically, we will make the following approximation. Firstly, the energy is dominated by the contributions near the defects, in a (geodesic) disk $D_j$ of some radius $R$ about the defects. Thus, the integral over the full domain is approximately the integral over the union of these disks. Secondly, we assume the defects are reasonably well-separated. Under this assumption, as well as the assumption that $R$ is not too large, we can approximate $\phi_k \approx \phi_{jk} + \rho_j\partial_{\rho_j} \phi_k $ on disk $D_j$, where $\phi_{jk} := \phi_k({\bf r}_j)$ is constant. Note that care must be taken to define the angles $\phi_{jk}$ properly, as described in Section~\ref{sec:orientation}. We will assume $R$ is small, and neglect curvature contributions, so that the area element on $D_j$ is approximately $\rho_j d\rho_j d\phi_j$.  

The first, constant part of this approximation of $\phi_k$ considers the defects to be `frozen in', and will result in terms that have no distance dependence. While this initially appears strange, as it seems to be an `infinite range' interaction, it is really an artefact of our choice of frame. It gives the relevant interactions when the defects are assumed to sit at umbilics, and the interaction terms it generates give pure torques, and align the defects relative to one another and to the principal curvature directions near the umbilics. The distance-dependent part of the approximation will result in interactions proportional to $\rho_{jk}^{-2}$ in the defect separation $\rho_{jk}$, as well as the angles between them. 

For simplicity, we will compute only the first, `frozen in' part of the interaction here. This already gives interesting physics, and a global alignment problem for the defects. With this in mind, we take $\phi_k \approx \phi_{jk}$ on $D_j$, and define the constant,
\begin{equation}
    \beta_j = \sum_{k\neq j} q_k \phi_{jk}.
\end{equation}
Loosely, $\beta_j$ is the angular momentum of the entire defect gas as seen by the defect at ${\bf r}_j$. Under this approximation we have
\begin{equation}
     \theta_0 \approx q_j\phi_j + \beta_j,
\end{equation}
on disk $D_j$. It follows that
\begin{equation}
    \cos(2\theta_0) \approx \cos \left(2q_j\phi_j + 2\beta_j\right),
\end{equation}
on $D_j$. Then we expand using standard trig identities, 
\begin{equation}
    \cos 2 \theta_0 \approx \cos(2q_j \phi_j)\cos\left(2\beta_j \right) -  \sin(2q_j \phi_j)\sin\left(2\beta_j \right)
\end{equation}
Now, we integrate this term over $D_j$. The integral splits into a radial and an angular part. For $q_j$ an integer or half integer, integrating $\sin( 2 q_j \phi_j)$ and $\cos( 2q_j \phi_j)$ over a full period gives zero. Here, we find a genuine difference for $p$-atics, where fractional charges would give a nonzero contribution from this integral. Even in the nematic case, we should be careful, as it is possible to have defects with $q_j=0$ in our system, when those defects sit exactly at umbilic points with the same winding. The $q_j=0$ contribution results in an energy contribution
\begin{equation}
    \pi R^2\cos(2\beta_j)\delta(q_j).
\end{equation}
It remains to consider the other term. By the same approximations, we find 
\begin{equation} \label{eq:ext_curvature_integrand}
    \theta_0 \sin(2\theta_0) \approx \left(q_j\phi_j + \beta_j\right)\left(\sin(2q_j \phi_j)\cos\left(2\beta_j \right) + \cos(2q_j\phi_j)\sin\left(2\beta_j \right) \right),
\end{equation}
on the disk $D_j$. When $q_j$ is a nonzero half integer, the term with the constant prefactor $\beta_j$ vanishes when integrated over the disk, by the same argument we just made for the other integral. Otherwise, when $q_j=0$ it yields a contribution
\begin{equation}
    \pi R^2 \beta_j\sin (2\beta_j) \delta(q_j)
\end{equation}
The remaining terms being integrated are proportional to $\phi_j \sin(2 q_j\phi_j)$ and  $\phi_j \cos(2 q_j\phi_j)$. The angular integral of the latter term vanishes, while the former is,
\begin{equation}
    \int_0^{2\pi} q_j\phi_j \sin(2 q_j\phi_j) d\phi_j = -4\pi q_j^2. 
\end{equation}
The radial part of the integral then gives a constant factor of $R^2/2$. Finally, we arrive at our expression for the energy,
\begin{equation}
    E = E_0  +\Delta \pi R^2 \sum_j \left( -2q_j^2 \cos (2\beta_j) +  \beta_j\sin (2\beta_j) \delta(q_j) +\frac{1}{2}\cos(2\beta_j)\delta(q_j) \right) + \text{self-energies} + O(\Delta^2).
\end{equation}
Here, we have introduced $\beta_j = \sum_{k\neq j}q_k\phi_{jk}$. The extra term in the energy yields a torque which sets relative angles between the defects, and depends on the distance between defects only implicitly through the fact we have introduced a cutoff radius $R$ for the integration such that the defect separation must be meaningfully larger than $R$. Distance-dependent terms come in from the linear part of $\phi_k$ which we have neglected: these scale like $q_jq_kR^3/\rho_{jk}^2$. In general this is a complex, nonlinear alignment problem, which warrants a detailed numerical study. There has been some attempt to study the effect of extrinsic curvature on defects in active nematics~\cite{pearceCouplingTopologicalDefect2022, pearceDefectOrderActive2020}, but to our knowledge no study has examined this alignment problem in detail.  

Here, we have not allowed for the possibility for spiral charge. We include this by inserting the additional terms proportional to $\log \rho_j$ into $\theta_0$. In the approximation where defects are widely separated, we fix a defect $j$ and approximate $\rho_k$ as constant and equal to the distance between the defects for each $k \neq j$. Carrying out the integrals in this case is similarly straightforward and results in terms dependent on both $\beta_j$ and $\alpha_j = \sum_{k\neq j}p_k\rho_{jk}$, where $\rho_{jk}$ is the geodesic distance between the two defects and $p_k$ is the spiral charge of defect $k$. 

A final remark: the reader should be careful that, throughout this section, we have written the director in the very specific form ${\bf n} = \cos \theta \, {\bf e}_1 + \sin \theta \, {\bf e}_2$ where ${\bf e}_1, {\bf e}_2$ are the principal curvature directions, not an arbitrary basis. Thus, the defect charges $q_j$ are windings relative to the zeros of this field. This should be taken into account when attempting to apply our results to a setting where the director is written in local coordinates on the surface. We have also assumed that the curvature anisotropy is both small and approximately constant. Without these assumptions evaluating the integrals is significantly more challenging and perhaps has to be done numerically---our results, however, give insight into the expected behaviour.

\subsection{Elastic Anisotropy}
Elastic anisotropy is known to be quite important in shaping equilibrium structures~\cite{lavrentovich_splaybend_2024}. Many previous works have discussed the effect of elastic anisotropy on local defects profiles and the resulting change to their interactions, and a detailed analysis was recently given by Houston~\cite{houston_role_2026}, who established the form of the interaction and showed that the linear order contribution coming from elastic anisotropy induces a separation-independent torque depending on the sign of the anisotropy and the relative orientation between the defects. 
We revisit this interaction from the geometric field theory perspective, using spiral charge to encode the orientation, and derive the linear order in $\epsilon$ correction to the force acting on the defects. 

The full intrinsic Frank energy with two elastic constants is, 
\begin{equation}
    E = \int_M \frac{K_1}{2} (\nabla \cdot {\bf n})^2 + \frac{K_3}{2} ({\bf n} \cdot \nabla {\bf n})^2 \mu.
\end{equation}
As usual, we write ${\bf n} = \cos \theta {\bf e}_x + \sin \theta {\bf e}_y$. We can replace ${\bf e}_x, {\bf e}_y$ with a general orthonormal frame if we are working on a curved surface, but for the remainder of this section we will consider anisotropic directors on a flat surface. Set ${\bf n}_\perp = -\sin\theta {\bf e}_x + \cos \theta {\bf e}_y$. We will define the mean elastic constant $\bar{K} = (K_1+K_3)/2$, and an anisotropy parameter $\epsilon = (K_3-K_1)(K_3+K_1)$. We work in units where $\bar{K} = 1$. Then we have~\cite{houston_role_2026},
\begin{equation}
    E = \frac{1}{2}\int_M |\nabla {\bf n}|^2 + \epsilon\left(|{\bf n} \cdot \nabla {\bf n}|^2- |{\bf n}_\perp \cdot \nabla {\bf n}_\perp|^2 \right)\mu.
\end{equation}
This particular form for the energy is the most instructive, as it depends on the inherent symmetries that exist in 2D. As was the case for extrinsic curvature, this energy cannot be written purely in terms of $A$, because it depends explicitly on ${\bf n}_\perp$ and not just the gradient. 

We insert $f = \left(|{\bf n} \cdot \nabla {\bf n}|^2- |{\bf n}_\perp \cdot \nabla {\bf n}_\perp|^2 \right)$ into the general formula Eq.~\eqref{eq:energy_expansion} given in Section~\ref{sec:perturbations}. We find that, 
\begin{equation} \label{eq:aniso_functions}
    \begin{aligned}
    f &= ((\partial_x \theta)^2-(\partial_y \theta)^2)\cos(2\theta) + 2\partial_x \theta \partial_y \theta \sin(2\theta), \\
        \partial_\theta f &= -4((\partial_x \theta)^2-(\partial_y \theta^2 ))\sin(2\theta) + 8\partial_x \theta \partial_y \theta \cos(2\theta), \\
        \nabla\theta_0 \cdot \partial_{\nabla \theta }f &= 2|({\bf n }\cdot \nabla \theta){\bf n}|^2 - 2 |({\bf n }_\perp\cdot \nabla \theta){\bf n}_\perp|^2
    \end{aligned}
\end{equation}
We see that $\nabla \theta \cdot \partial_{\nabla \theta }f = 2f$, simplifying the calculations. Moreover, $\partial_\theta f$ is very similar to $f$, simply with the roles of $\cos(2\theta)$ and $\sin(2\theta)$ interchanged. Using our general expansion for the energy, Eq.~\eqref{eq:energy_expansion}, we see we need to compute the integral of $f(\theta_0, \nabla \theta_0)$ and also $\theta_0 \partial_\theta f(\theta_0, \nabla \theta_0)$. 

Let us first compute the integral of $f$. We can write, 
\begin{equation}
    f =  \sum_{j,k} q_jq_k\left( (\partial_x \phi_j \partial_x\phi_k - \partial_y \phi_j \partial_y\phi_k)\cos(2\theta) + 2(\partial_x \phi_j \partial_y\phi_k + \partial_x \phi_k \partial_y\phi_j)\sin(2\theta) \right).
\end{equation}
We note that this expression, much like the situation for extrinsic curvature, involves many-body interactions---we have a sum over pairs of defects, and then within $\theta$ we have an additional sum over all defects. We then find that, 
\begin{equation}
    \begin{aligned}
        (\partial_x \phi_j \partial_x\phi_k - \partial_y \phi_j \partial_y\phi_k) &= -\frac{\cos(\phi_j+\phi_k)}{\rho_j \rho_k} ,\\
        (\partial_x \phi_j \partial_y\phi_k + \partial_x \phi_k \partial_y\phi_j) &= -\frac{\sin(\phi_j+\phi_k)}{\rho_j \rho_k}. 
    \end{aligned}
\end{equation}
Now we can go through and apply standard trig identities, as well as the definition of $\theta_0$, to finally obtain 
\begin{equation}
    f =  -\sum_{j,k} \frac{q_jq_k}{\rho_j \rho_k}\cos \left(\phi_j + \phi_k - 2\sum_l q_l \phi_l  \right).
\end{equation}
As we did with extrinsic curvature, we decompose the integral of $f$ into integrals over disks $D_j$ about each of the defects ${\bf r}_j$, and we assume that the defects are widely separated so that $\phi_k \approx \phi_{jk}$ and $\rho_k \approx\rho_{jk}$ (the geodesic distance between defects) are approximately constant on each $D_j$---the next order contribution will lead to terms which decay faster with separation, and we will neglect these as we did for curvature anisotropy. The we perform the angular integration over $D_j$. 

There are two cases, where $j=k$ and when $j \neq k$. Let us first examine the $j=k$ case, which gives an energy contribution which does not depend on distance. In this case, the integrand is 
\begin{equation}
    f =  -\frac{q_j^2}{\rho_j^2 }\cos \left(2\beta_j \right).
\end{equation}
The angular integration gives a constant $2\pi$. Computing the radial integral over $\rho_j$ then gives a constant $\log(R/r_\text{core})$ depending on a core cutoff and the size $R$ of the disk $D_j$, which should be chosen to be significantly smaller than the defect separation while also being larger than the core size. Thus, the contribution is 
\begin{equation}
  -\frac{\pi}{2}\log(R/r_\text{core}) \cos \left(2\beta_j \right).
\end{equation}
When $j\neq k$, the result of the integration depends on the sign of $q_j$. If $q_j = -1/2$, the integral vanishes due to the oscillatory nature of $\cos$. If $q_j = +1/2$, the integrand then becomes a constant, independent of $\phi_j$ and we have, 
\begin{equation}
    \int_0^{2\pi} f d\phi_j \approx -\frac{2\pi}{\rho_j}\sum_{k \neq j} \frac{q_k}{\rho_{jk}}\cos \left( \phi_{jk}-\sum_{l \neq j} q_l \phi_{jl}\right)\delta(q_j - 1/2). 
\end{equation}
The dependence of this result on $q_j$ is expected: anisotropy contributes when there is a dipolar symmetry breaking, which is the case for a $+1/2$ defect but not a $-1/2$ defect.  The energy contributes a $1/r$ integration between the defects $j, k$ as long as at least one of these defects has winding $+1/2$, and this interaction depends on the angles between all defects in the system, and is therefore a many-body effect in the case that there are more than two defects. It therefore induces torques on all the defects, provided at least one has winding $+1/2$. This latter condition is a necessity on a sphere or torus or infinite plane with uniform boundary conditions, but need not be the case on a handlebody of genus larger than 1, for example.  

The next contribution is from $\theta_0\partial_\theta f(\theta_0, \nabla \theta_0) $. A short calculation, almost identical to the calculation we did for $f$, shows that, 
\begin{equation}
        \partial_\theta f = -4\sum_{j,k} \frac{q_jq_k}{\rho_j \rho_k}\sin \left(\phi_j + \phi_k - 2\sum_l q_l \phi_l  \right). 
\end{equation}
Continuing to employ the same approximation, on a disk $D_j$ we find that,
\begin{equation}
    \theta_0 \partial_\theta f \approx -4\sum_j\left(q_j\phi_j +\beta_j\right) \sum_{k,l} \frac{q_kq_l }{\rho_{k}\rho_{l}}\sin \left(\phi_k + \phi_l - 2\sum_n q_n \phi_{jn}  \right).
\end{equation}
The angular integral of this is analogous to many integrals we have done before, and depends on the values of $k,l$. When $k,l = j$, we again obtain a term which is independent of the defect separations,  
\begin{equation}
    -2\pi(q_j\pi+\beta_j)\log(R/r_\text{core}) \sin \left(2\beta_j \right).
\end{equation}
Now we consider the remaining cases, where at least one of $k, l$ is not equal to $j$. Firstly, focus on the term with the constant prefactor $\sum_{m \neq j}q_m\phi_{jm}$. This contributes only when exactly one of $k, l$ equals $j$, and then only for $q_j = 1/2$, exactly as for the integral of $f$. Its contribution is,  
\begin{equation}
   -\frac{8\pi}{\rho_j} \left(\sum_{m \neq j} q_m \phi_{jm} \right) \sum_{k \neq j} \frac{q_k }{\rho_{jk}} \sin \left( \phi_{jk}-\sum_{l \neq j} q_k \phi_{jl}\right)\delta(q_j - 1/2).
\end{equation}
We picked up an extra factor of two here because we have contributions from both $k=j$ and $l=j$ in the integrand. 

The other part, with the prefactor $q_j\phi_j$, is entirely analogous to integrals we carried out in the previous section on extrinsic curvature. It contributes a nonzero energy regardless of the values of of $k, l$, and also $q_j$. However, to simplify matters slightly, we note that when neither of $k,l$ is equal to $j$, the integral scales like $1/(\rho_{jk}\rho_{kl})$, i.e., it effectively depends on the square of the separation between defects $j$ and $k$ in the case $k=l$, and on the area of the triangle made by the three defects $j,k,l$ when $k\neq l$. Both of these are shorter range interactions than the $1/r$ interaction we have already found. We will therefore neglect these terms, although they are straightforward to calculate if the reader is interested in doing so. 

The remaining case then boils down to the angular integral of 
\begin{equation}
    -8 q_j\phi_j \sum_{k \neq j} \frac{q_jq_k }{\rho_{j}\rho_{jk}}\sin \left((1-2q_j)\phi_j + \phi_{jk} - 2\sum_{l \neq j} q_l \phi_{jl}  \right),
\end{equation}
where we again picked up an extra factor of two because we have contributions from both $k=j$ and $l=j$. Using the standard formula
\begin{equation}
    \int_0^{2\pi} x\sin(nx + C) = -\frac{2\pi}{n}\cos(C),
\end{equation}
we find the contribution from this term is 
\begin{equation}
    \frac{4\pi}{1-2q_j} \sum_{k \neq j} \frac{q_k}{\rho_{j}\rho_{jk}}\cos \left(\phi_{jk} - 2\sum_{l \neq j} q_l \phi_{jl}  \right).
\end{equation}

The final contribution comes from the integral of $\nabla\theta_0 \cdot \partial_{\nabla \theta }f$, but as we have observed, this is just $2f$, and so we have already calculated this. 

It remains to consider the two boundary terms in Eq.~\eqref{eq:energy_expansion}. For extrinsic curvature, both of these contributions vanished. Examining Eq.~\eqref{eq:aniso_functions}, we see first that $\nabla\theta_0 \cdot \partial_{\nabla \theta }f \neq 0$, and also that the source term for the linear correction $\theta_1$ contains derivatives of $\theta_0$, which means it can pick up distributional singularities. Both integrals may therefore be nonzero. However, since the first of these terms involves $df$, it decays at order $1/r^2$ in distance from the defects, rather than $1/r$. Similarly, as $\theta_1$ involves a convolution of a source that goes like $1/r$ with a Green's function, the term that involves integrating $\theta_0 \star d\theta_1$ also decays faster than $1/r$. Moreover, as these are boundary integrals we never integrate out radial factors, so there cannot be a distance-independent term coming from them. Thus, we choose to neglect both of these terms---since we are assuming the defects are well-separated, these shorter range interactions are dominated by the terms we have already computed. 

Putting all this together, we have computed two contributions to the energy: those that are independent of distance, and those that scale like $1/r$ in defect separation. The longest range part is the former, which gives us 
\begin{equation}
    E \approx E_0 + \epsilon\pi c\left( \frac{1}{4} \cos(2\beta_j) + (q_j\pi+\beta_j) \sin \left(2\beta_j \right) \right) + O(\epsilon^2, 1/r),
\end{equation}
where $E_0$ is the interaction energy when $\epsilon =0$. Here, we again use $\beta _j = \sum_{l \neq j} q_l \phi_{jl}$ to simplify notation, and we have written $c = \log (r_\text{core}/R)$ for the constant which depends on the core radius and the scale of the disks we integrate over. We emphasise that this formula is valid on a curved surface provided one uses the appropriate geodesic distances and angles. The contributions from non-zero spiral charge are readily computed using similar formulas to those we have calculated, and similarly result in global torques. 

Comparing this with the formula derived by Houston~\cite{houston_role_2026}, we find some key differences due to our differing approaches. Houston considers the linear in $\epsilon$ correction arising from only pairwise interactions, and is able to perform his integral in bipolar coordinates, bypassing the need for approximations and hence the dependence on defect core size that arises when considering arbitrary numbers of defects. Additionally, \cite{houston_role_2026} uses only the $\theta_0$ solution and not the $\theta_1$ correction as we do, and this results in missing the cross terms which produce the $\sin(2\beta_j)$ term in our result. However, the core fact that elastic anisotropy induces distance-dependent torques at linear order, proportional to the angles between them, remains consistent across both works. 

It is also worth comparing this with extrinsic curvature. As with extrinsic curvature, the key feature of the interaction is the sum of angles $\beta_j$ which introduces a global, distance-independent torque on the defects. We also have additional three-body interactions, proportional to charge combinations $q_j q_k q_l$ and to the area of the triangle made by the defects, which we have not examined. The energy is optimised via a complex nonlinear alignment problem which, in general, does not seem straightforward to solve. As for extrinsic curvature, a detailed numerical analysis of the effect of elastic anisotropy on defect position and alignment would be a helpful complement to our calculation.

\section{Discussion}\label{sec:discussion}
In this work, we have used geometric field theory as a unifying language for the systematic understanding of the particlelike kinematics of topological defects in passive 2D nematic materials. The approach, we argue, lays bare the origin and nature of various contributions to defect interactions.
The basic component of such a picture is the Green's function of the Laplacian, which captures how the interaction changes when (geodesic) distances between points on the domain are altered, either via changes in global topology or intrinsic geometry. Further effects, we show, are a consequence of the gauge \textit{non}invariance of the elastic free energy. Using Hodge theory, we codify this via harmonic `excitations' of the global ground state. Different excitations can be thought of as energy minimisers under the imposition of different constraints; a gauge invariant theory would be indifferent to such constraints. The approach is shown to capture the effects of both topological solitons and relative defect orientations. We also examine perturbations to the form of the energy that would otherwise be forbidden in a gauge invariant construction. These result from intrinsic curvature and elastic anisotropy. In both, such modifications lead to non-linearities in the corresponding Euler--Lagrange equations, giving rise to \emph{induced} many-body interactions between defects. This is despite the underlying symmetry group being abelian, a fact that would ensure that interactions between defects are only pairwise in a gauge invariant theory.

In addition to conceptual insight, the approach has several other advantages.

First, the framework provides a convenient structure for several calculations. Many of the integrals we need to perform are very straightforward, arising from the $L^2$ inner product, and they frequently vanish or localise on the singularities due to orthogonality. In most cases, to obtain the field we only need to solve a linear, Poisson-type problem, which can be done using the Green's function for the Laplacian. We see how this Green's function essentially controls all the interactions in the system. The Green's functions can be found in practice for a wide number of relevant cases using standard techniques such as conformal mapping or image constructions~\cite{vitelli_defect_2004, turner_vortices_2010, vromansOrientationalPropertiesNematic2016, tangOrientationTopologicalDefects2017}.

Second, problems involving a nonlinear contribution to the Euler--Lagrange equation, such as extrinsic curvature of the surface or elastic anisotropy in the free energy, can be addressed perturbatively; it is relatively easy to determine approximate, closed form expressions from the linear order contributions to the energy. Nevertheless, these involve complex many-body interactions, and the optimal position and alignment of defects is difficult to obtain in the general case. Numerical work in this area would therefore be highly instructive.

Third, closely related geometric constructions can, in principle, be extended to 3D by changing the symmetry group from $U(1)$ to $SU(2)$. Although the resulting theory is substantially richer in 3D due to its non-abelian nature, the framework of geometric field theory nevertheless helps to translate concepts across dimensions in a way that, for example, complex analysis approaches do not. The present strategy also extends to polar materials or general $p$-atics. The only difference here is the numerical value of the defect charge, which will only give a substantial modification to the elastic/curvature anisotropy contributions to the energy, as certain terms will not vanish for $p \neq 2$. 

There are also several areas that require further work. We have not discussed hydrodynamics, which of course plays a fundamental role even for a passive liquid crystal. A natural extension of this work would be to account for hydrodynamics, both passive and active, and to try to derive the interactions between defects in the presence of fluid flows, as well as the interaction between nematic defects and fluid vortices. Hydrodynamics can be dealt with using the same framework of geometric field theory we have employed here, with two fields encoding the nematic director and the velocity, reduced to a `defect-vortex gas'. In addition to the energy functional, we must also include an appropriate dissipation functional for the nematic as in Refs.~\cite{tangTheoryDefectMotion2019, zhang_dynamics_2020}, with the dynamics being given by the minimisation of the sum of the energy and dissipation. 

Another important extension is to treat a nematic on a deformable surface~\cite{al-izziActiveFlowsDeformable2021,al-izziMorphodynamicsActiveNematic2023}. This can be accomplished by taking the energy which contains both the intrinsic and extrinsic contributions, and supplementing this by additional terms representing the energy of the surface itself. The dynamics of the nematic can be reduced to a defect gas picture as we have done here, while the dynamics of the surface can be dealt with by treating the metric and second fundamental form as dynamical variables~\cite{morrisActiveMorphogenesisEpithelial2019}. Our calculations here can be applied to this case, and it is fairly easy to treat situations where the evolution of the surface is constrained to be a conformal map, since the conformal factor neatly determines both the Green's function and the geometric potential. Since nematics are now frequently employed as models for tissues and membranes, and defects within the nematic field are well known to play an important role in morphogenesis, being able to analytically address questions of how deformable surfaces respond to the presence of a nematic defect and how this feeds back into the force acting on the defect would be highly relevant for fundamental biological questions.

\begin{acknowledgements}
RGM acknowledges the EMBL Australia program and funding from the Australian Research Council Centre of Excellence for Mathematical Analysis of Cellular Systems (CE230100001).
\end{acknowledgements}

\bibliography{references.bib}

\end{document}